\documentstyle[12pt]{article}
\def\prl{Phys.\ Rev.\ Lett.\ }

\def\prb{Phys.\ Rev.\ B }
\def\neel{N\'{e}el }
\begin{document}
\setlength{\baselineskip}{.25in}
\newcommand{\tb}{\tilde{b}}
\newcommand{\tm}{\tilde{m}}
\newcommand{\cj}{{\cal J}}
\newcommand{\etp}{\eta_{p}}
\newcommand{\mt}{\mbox{ (mod 2) }}
\newcommand{\bqo}{\bar{Q}_1}
\newcommand{\hx}{\hat{x}}
\newcommand{\hy}{\hat{y}}
\newcommand{\cH}{{\cal H}}
\newcommand{\bN}{{\bf n}}
\newcommand{\tn}{\tilde{n}}
\newcommand{\bL}{{\bf L}}
\newcommand{\bb}{\bar{b}}
\newcommand{\hs}{\hat{S}}
\newcommand{\hS}{\hat{S}}
\newcommand{\br}{{\bf r}}
\newcommand{\bR}{{\bf R}}
\newcommand{\mq}{\bar{Q}}
\newcommand{\blam}{\bar{\lambda}}
\newcommand{\bk}{{\bf k}}
\newcommand{\gk}{\gamma_{\bk}}
\newcommand{\ok}{\omega_{\bk}}
\newcommand{\half}{\mbox{$\frac{1}{2}$}}
\newcommand{\hmu}{\hat{\eta}}
\newcommand{\bc}{\bar{c}}
\newcommand{\bn}{{\bf N}}
\newcommand{\ttau}{\tilde{\tau}}
\newcommand{\tg}{\tilde{g}}
\newcommand{\bs}{{\bf S}}
\newcommand{\hbs}{\hat{\tilde{S}}}
\newcommand{\trho}{\tilde{\rho_s}}
\newcommand{\si}{\hs_{1}^{2}}
\newcommand{\sic}{\hs_{2}^{1}}
\newcommand{\qi}{q}
\newcommand{\qc}{q^{\ast}}
\newcommand{\wn}{\omega_n}
\newcommand{\mnb}{n_b ( \mbox{mod}\, 4 )}
\newcommand{\ms}{2S ( \mbox{mod}\, 4 )}
\newcommand{\mst}{2S ( \mbox{mod}\, 2 )}
\newcommand{\bch}{\bar{\chi}}
\newcommand{\hU}{\hat{U}}
\newcommand{\hV}{\hat{V}}
\newcommand{\crl}{\{ n_{\ell} \} }
\newcommand{\nl}{n_{\ell}}
\newcommand{\cab}{\chi_{b}^{a}}
\newcommand{\cba}{\chi_{a}^{b}}
\newcommand{\caa}{\chi_{a}^{a}}
\newcommand{\hT}{\tilde{T}}
\newcommand{\bG}{{\bf G}}
\newcommand{\mf}{\mbox{ (mod 4) }}
\renewcommand{\topfraction}{0.95}
\renewcommand{\textfraction}{0.05}
\vspace{1in}
\begin{center}
\bf
QUANTUM ANTIFERROMAGNETS IN TWO DIMENSIONS\\
\vspace{0.75in}
\normalsize\rm
Subir Sachdev \\
Department of Physics\\
Yale University\\
P.O. Box 2157, Yale Station\\
New Haven CT 06520, U.S.A.\\
\vspace{0.3125in}
\tt
Lectures presented at the summer course on \\
``Low Dimensional Quantum
Field Theories for Condensed Matter Physicists'',\\
24 Aug. to 4 Sep. 1992,
Trieste, Italy\\
\rm
\vspace{0.2in}
\end{center}
\setlength{\baselineskip}{.25in}
\rm
\begin{center}
ABSTRACT
\end{center}
I review recent work, performed in collaboration primarily with N. Read
and Jinwu Ye, on the properties of quantum antiferromagnets in two
dimensions. The emphasis is on the properties of the antiferromagnet
in states which do not have any long-range magnetic order. The
universal spin dynamics in the quantum critical region of
number of frustrated and random antiferromagnets is studied;
implications for neutron scattering
experiments in the lightly-doped cuprates are noted.
The nature of the quantum-disordered phase of non-random frustrated
antiferromagnets is examined in some detail: the states found have
({\em i\/})
collinear spin correlations, spin-Peierls or valence-bond-solid order,
and confined spinons,
order and confined spinons or ({\em ii\/}) coplanar spin correlations,
no spin-Peierls order and deconfined bosonic spinons.

\section*{\normalsize\bf 1. INTRODUCTION}
The study of quantum antiferromagnets goes back to very early work
of Bethe~\cite{bethe}, Anderson~\cite{philold} and Kubo~\cite{kubo}.
However the subject has recently seen a tremendous amount of renewed
theoretical and experimental interest.
We will restrict the discussion here
to antiferromagnets described by the following simple
Hamiltonian:
\begin{equation}
{\cal H} = \sum_{i,j} J_{ij} \hat{\bf S}_i \cdot \hat{\bf S}_j
\label{hamil}
\end{equation}
where the $\hat{\bf S}_i$ are spin $S$
quantum spin operators on the sites, $i$, of
a $2$-dimensional lattice. The antiferromagnetic exchange constants,
$J_{ij} > 0$, are short-ranged and possibly random.
The early work~\cite{bethe,philold,kubo} focused
on the nature of the ground states of ${\cal H}$ with magnetic
long-range-order (LRO):
these are states in which there is a spin-condensate:
\begin{equation}
\langle {\bf S}_i \rangle = {\bf m}_i \neq 0
\end{equation}
The situation changed dramatically with the
suggestion by Anderson~\cite{anderson} in 1987 that
quantum disordered ground states of two dimensional
quantum antiferromagnets  are relevant
to the phenomenon of high temperature
superconductivity in the cuprates~\cite{nobel}.
The magnetically disordered states are those in which
\begin{equation}
\langle {\bf S}_i \rangle = 0
\label{qdp}
\end{equation}
We emphasize that the spin-glass ground state is not of this type; rather
it is a magnetically ordered ground state in which ${\bf m}_i$ is a
random function of $i$.
The following questions arise immediately in the study of any quantum
disordered phase:
\begin{itemize}
\item
Is the quantum-disordered ground state characterized completely by
Eqn. (\ref{qdp}) ? Or are there
additional broken symmetries that must be present ? Naively, one may
conclude that the first of these is the most natural possibility.
Many investigators have addressed this issue and found however that
other broken symmetries appear more often than was initially expected.
Only under special conditions, which N. Read and I delineated in
Refs.~\cite{self4,self5,kagome}, does a featureless spin-fluid state
appear. This work will be reviewed in Section 6.
\item
What is the nature of the excitations above the ground state ? Do the
individual quanta of excitations carry fractional spin quantum numbers
(S=1/2) ? Such excitations are conventionally called spinons. A second
possibility is that the spinons are always confined in pairs, leading
to excitations which have only integral spin.
\item
If the spinons are unconfined, what are their statistics ?
In non-random antiferromagnets, the spinons are usually
separated from the ground state by a gap, and physical conditions with a
sufficiently dilute concentration of spinons can always be defined to make
the question of their statistics meaningful.
\end{itemize}
Most of the questions above will be addressed in the course of these
lectures. We will use a combination of semiclassical and and a number
of different large-$N$ expansions to address these issues. In
particular, much will
be learned by comparing and reconciling the results of
the different expansions.

There are several motivations for addressing these and
other related questions:
\begin{itemize}
\item
The cuprates undergo a transition to a quantum disordered ground state
upon doping with a small concentration of mobile holes. In $La_{2-x}
Sr_x Cu O_4$ this occurs somewhere between $x=0.02$ and $x=0.07$.
At lower dopings the holes are localized and a Hamiltonian as in
(\ref{hamil}) is adequate.
At larger dopings the holes become mobile,
requiring study of quantum-disordered
antiferromagnets in the presence of propagating holes
A number of neutron
scattering experiments have given
us a detailed picture of the spin
dynamics of these
antiferromagnets~\cite{hayden1,cheong,keimer,hayden2,mason,thurston,keimer2,matsuda},
much of which is poorly understood.
\item
The spin-1/2 antiferromagnet on a kagom\'{e} lattice may posses a quantum
disordered phase. Recently there have been a number of experimental
realizations of Heisenberg antiferromagnets on a kagom\'{e} lattice:
the layered kagom\`{e} antiferromagnets
$Sr\-Cr_{8-x}\-Ga_{4+x}\-O_{19}$,
$K Cr_3 (OH)_6 (SO_4 )_2$~\cite{ramirez,broholm,broholm2}
and layers of ${}^3 He$ on graphite~\cite{greywall,elser}.
\end{itemize}

As a noted above, there have been a number of detailed dynamic
neutron scattering experiments
on lightly-doped
cuprates~\cite{hayden1,cheong,keimer,hayden2,mason,thurston,keimer2,matsuda}
and layered frustrated
antiferromagnets~\cite{greywall,elser,ramirez,broholm,broholm2}.
One feature of the dynamic neutron scattering on the lightly-doped
cuprates ($La_{2-x} Sr_x Cu O_4$ for $0.02 < x < 0.05$)
that is
particularly intriguing
is that
the overall frequency scale of the spin excitation spectrum appears to be given
simply by
the absolute temperature. In particular, it appears to be
independent of all microscopic energy scales
{\em e.g.\/} an antiferromagnetic exchange constant.

We have recently proposed~\cite{jinwu} that
this anomalous dynamics
is a very general property of finite $T$,
spin fluctuations in the \underline{\sl quantum-critical}~\cite{sudip}
region.
This is the region in the
temperature-coupling constant plane where the spin fluctuations are
controlled by the critical fixed point between the magnetically
ordered and the quantum disordered phase. The
antiferromagnet notices the thermal deviation from $T=0$ at a scale
\underline{shorter} than that at
which it notices that the coupling constants are not
exactly critical. The system is neither
fully ordered or fully disordered at long distances, but remains in
limbo controlled by the critical point between them; the dynamics
exhibits features of both phases and the energy scale is set solely
by the absolute temperature.
In our work~\cite{jinwu}
universal scaling functions and exponents for the
quantum-critical dynamics of frustrated, doped, and random
antiferromagnets have been proposed.
Complete scaling functions and
exponents have been calculated for
a model system - non-random, frustrated
two-dimensional Heisenberg antiferromagnets with a vector
order-parameter. These results will be reviewed in Sections 4 and 5 and
represent the first calculation of
dynamic scaling functions in a two-dimensional quantum phase
transition.

Most of the work reviewed here will be on non-random two-dimensional
antiferromagnets. There will however be a limited discussion of the
consequences of randomness in Section 5. For the most part, random
quantum antiferromagnets are only very poorly understood; there has however
been some recent progress~\cite{qsg,qlrm} for which the reader is invited
to consult the original papers.

We will begin this review with a derivation of the coherent-state,
path-integral
formulation of quantum antiferromagnets. The representation of the
semiclassical fluctuations of antiferromagnets by the non-linear sigma
model will be discussed in Section 3. In Section 4,
the properties of the $O(M)$ non-linear
sigma model will be studied in a large=$M$ theory, with a particularly
emphasis on the quantum-critical region. This will lead to a more
general discussion of quantum-criticality in Section 5. Finally the
quantum disordered state will be discussed in Section 6.

\section*{\normalsize\bf 2. COHERENT STATE PATH INTEGRAL}

In this section we shall present a derivation of a coherent state path
integral formulation of the dynamics of a generic  antiferromagnetic.
There will be no approximations made and the results
are essentially exact. As identical manipulations have to be performed
on all the spins of the antiferromagnet, we will drop the site index
in the following and simply present the path integral formulation for
a single spin. The path integral for the lattice problem will
involve direct copies of the single site result.
The results of this section are ``well-known'', although the
formulation here was first discussed in Ref.~\cite{self1}; it has the
advantage of explicitly preserving spin-rotation invariance.

As a first step, we need to define the spin coherent states
The reader is urged at this point to consult the very instructive
discussion on the properties and uses of coherent states in general in
the introduction to the book edited by Klauder~\cite{jrk}.
We restrict our discussion here to the case of $SU(2)$ spins.
The conventional complete basis for the spin states at each site is the usual
one corresponding to the eigenstates of $\hat{\bf S}^2$ and
$\hat{S}_z$:
\begin{equation}
| S , m \rangle~~~~\mbox{with $m=-S \ldots S$}
\label{szstates}
\end{equation}
In the path-integral formulation it is more convenient to use instead
an overcomplete basis of states $| \bn \rangle$ labeled by the
points $\bn$ on the surface of the unit sphere. These states
are normalized, $\langle \bn | \bn \rangle =1$ but are not
mutually orthogonal.
Their crucial properties are that the expectation value of the spin
is given by
\begin{equation}
\langle \bn | \hat{\bf S} | \bn \rangle = S \bn,
\label{cohexp}
\end{equation}
and the completeness relation
\begin{equation}
\int \frac{ d \bn}{2 \pi} | \bn \rangle \langle \bn | = 1,
\label{complete}
\end{equation}
where the integral is over the unit sphere.
For $\bn = (0,0,1)$, the state $|\bn \rangle$ is easy to
determine; we have
\begin{equation}
|\bn = (0,0,1) \rangle = | S, m=S \rangle \equiv |\Psi_0 \rangle
\label{expect}
\end{equation}
We have defined this particular coherent state as a reference state
$|\Psi_0\rangle $ as it will be needed frequently in the following.
For other values of $\bn$ we can obtain $|\bn \rangle$ by a $SU(2)$
rotation of $|\Psi_0 \rangle$. It is not difficult to show in this
manner that
\begin{equation}
|\bn \rangle = \exp \left( z \hat{S}_{+} - z^{\ast} \hat{S}_{-} \right) |
\Psi_0 \rangle
\label{cohdef}
\end{equation}
where the relationship between the complex number $z$ and $\bn$ is
simplest in spherical co-ordinates:
\begin{eqnarray}
\bn &=& (\sin \theta \cos \phi , \sin \theta \sin \phi , \cos \theta ) \\
z &=& - \frac{\theta}{2} \exp ( - i \phi )
\label{nzdef}
\end{eqnarray}
The validity of the Eqn.~(\ref{expect}) determining the expectation
value of $\hat{\bf S}$ can be verified in a straightforward manner.

It will be useful for our subsequent formulation to rewrite the above
results in a somewhat different manner, making the $SU(2)$ symmetry
more manifest. Define the $2 \times 2$ matrix of operators $\hat{\cal S}$
by
\begin{equation}
\hat{\cal S} =  \left ( \begin{array}{cc} \hS_z & \hS_x - i\hS_y \\
\hS_x + i\hS_y & -\hS_z \end{array} \right ) .
\label{matop}
\end{equation}
Then Eqn.~(\ref{expect}) can be rewritten as
\begin{equation}
\langle \bn | \hat{\cal S}_{\alpha\beta} | \bn \rangle = S
Q_{\alpha\beta} ,
\label{cohprop}
\end{equation}
where the matrix $Q$ is
\begin{equation}
Q = \left ( \begin{array}{cc} N_z & N_x - iN_y \\
N_x + iN_y & -N_z \end{array} \right ) \equiv \bn \cdot \vec{\sigma}
\label{adjoint}
\end{equation}
where $\vec{\sigma}$ are the Pauli matrices.
Furthermore there is a simple relationship between $Q$ and the complex
number $z$. In particular, if we use the spin-1/2 version of the
operator in Eqn.~(\ref{cohdef})
\begin{equation}
U = \exp \left [ \left ( \begin{array}{cc}
0 & z \\
-z^{\ast} & 0
\end{array} \right ) \right ]
\end{equation}
($U$ is thus a $2\times 2$ matrix),
then we find
\begin{equation}
Q = U \sigma_z U^{\dagger}
\label{uq}
\end{equation}

We now proceed to the derivation of the coherent state path integral
for the partition function
\begin{equation}
Z = \mbox{Tr} \exp ( - \beta {\cal H} ( \hat{\bf S} ))
\label{canonical}
\end{equation}
where we have emphasized that the Hamiltonian ${\cal H}$ is a function
of the spin operator $\hat{\bf S}$; we will restrict the following
discussion to Hamiltonians in which ${\cal H}$ is a linear function of
any given $\hat{\bf S}$ on a fixed site. The ${\cal H}$ in
Eqn.~(\ref{hamil}) is certainly of this type.
The transformation of $Z$ into a path-integral begins with a standard
procedure: we will omit the details of some steps, referring the reader to the
introductory article in the book edited by Klauder~\cite{jrk}.
Briefly, the exponential in Eqn.~(\ref{canonical}) is written as the Trotter
product of a large number of exponentials each evolving the system
over an infinitesimal imaginary-time interval $\Delta \tau$; the
identity (\ref{complete}) is inserted between all the exponentials
and the matrix elements evaluated using (\ref{cohexp}). This yields
the result
may now be used to obtain
the following representation for the partition function
\begin{equation}
Z = \int {\cal D} \bn (\tau ) \exp \left\{ \int_0^{\beta} d\tau \left [
\langle \bn (\tau) | \frac{d}{d\tau} | \bn (\tau )\rangle
- \cH ( S\bn (\tau )) \right ] \right\} ,
\end{equation}
where $\cH (S \bn )$ is obtained by replacing every occurrence of
$\hat{\bf S}$
in the Hamiltonian by $S \bn $.
The first term in the action is the Berry phase term, $S_{B}$, and
represents the overlap between the coherent states at two
infinitesimally separated times. It can be shown straightforwardly
from the normalization condition $\langle \bn | \bn \rangle = 1$ that
$S_B$ must be pure imaginary. In the remainder of this section we
will manipulate $S_B$ into a physically more transparent form using
the expressions above for the coherent states.

Clearly, the $\tau$-dependence of $\bn (\tau )$ implies a $\tau$
dependent $z (\tau )$ through (\ref{nzdef}). From (\ref{cohdef}) we
have therefore
\begin{equation}
\frac{d}{d\tau} | \bn (\tau) \rangle = \frac{d}{d\tau}
\exp \left( z (\tau ) \hat{S}_{+} - z^{\ast} (\tau) \hat{S}_{-} \right) |
\Psi_0 \rangle
\label{dndt}
\end{equation}
Taking this derivative is however not so simple: I remind you that if
an operator $M$ does not commute with its derivative $dM/d\tau$ then
\begin{equation}
\frac{d}{d \tau } \exp (M) \neq \frac{d M}{d \tau } \exp (M)
\end{equation}
A careful analysis in fact leads to the general result~\cite{feynman}
\begin{equation}
\frac{d}{d\tau} \exp(M) = \int_0^1 du \exp(M(1-u)) \frac{dM}{d\tau} \exp
(Mu ) ,
\label{feynman}
\end{equation}
where $u$ is just a dummy integration variable.
Using (\ref{dndt}) and (\ref{feynman}) we find
\begin{eqnarray}
S_B &=& \int_{0}^{\beta} d \tau
\langle \bn (\tau )| \frac{d}{d\tau} | \bn (\tau) \rangle \nonumber \\
&=& \int_0^{\beta} d\tau
\int_0^1 du \langle \bn (\tau , u ) | \left ( \frac{\partial
z}{\partial \tau}  \hS_{+} - \frac{\partial
z^{\ast}}{\partial \tau} \hS_{-} \right ) | \bn (\tau , u ) \rangle
\end{eqnarray}
where $\bn (\tau , u )$ is defined by
\begin{equation}
| \bn (\tau , u) \rangle = \exp \left( u \left(
z (\tau ) \hat{S}_{+} - z^{\ast} (\tau) \hat{S}_{-} \right) \right) |
\Psi_0 \rangle
\end{equation}
 From this definition, three important properties of $\bn ( \tau , u)$
are immediately apparent
\begin{eqnarray}
\bn (\tau , u=1) &\equiv& \bn (\tau ) \nonumber \\
\bn (\tau , u=0) &=& (0,0,1)
\end{eqnarray}
\begin{displaymath}
\mbox{$\bn (\tau , u)$ moves with $u$ along the great circle between
$\bn (\tau , u=0)$ and $\bn (\tau , u=1)$}
\end{displaymath}
We can visualize the dependence on $u$ by imagining a string
connecting the physical value of $\bn (\tau ) = \bn (\tau , u=1)$ to
the North pole, along which $u$ decreases to 0.
We can also define a $u$-dependent $Q(\tau , u)$ from
Eqn.~(\ref{adjoint}); furthermore, if we choose
\begin{equation}
U (\tau , u) = \exp \left [ u \left ( \begin{array}{cc}
0 & z \\
-z^{\ast} & 0
\end{array} \right ) \right ]
\label{uu}
\end{equation}
then the relationship (\ref{uq}) remains valid for all $u$.
Now we use the expression (\ref{cohprop}) for the expectation value
of $\hat{\bf S}$ in any coherent state to obtain
\begin{equation}
S_B = S \int_0^{\beta} d\tau
\int_0^1 du \left [ \frac{\partial z}{\partial\tau}
Q_{21} (\tau , u )
- \frac{\partial z^{\ast}}{\partial\tau} Q_{12} (\tau ,u)
\right ] ,
\label{berry}
\end{equation}
As everything is a periodic function of $\tau$, we may freely
integrate this expression by parts and obtain
\begin{equation}
S_B = -S \int_0^{\beta} d\tau \int_0^{1} du \mbox{Tr} \left [
\left ( \begin{array}{cc}
0 & z(\tau) \\
-z^{\ast}(\tau ) & 0
\end{array} \right ) \partial_{\tau} Q(\tau ,u) \right ] .
\label{simple}
\end{equation}
where the trace is over the $2 \times 2$ matrix indices.
The definitions (\ref{uq}) and (\ref{uu}) can be used to easily
establish the identity
\begin{equation}
\left ( \begin{array}{cc}
0 & z(\tau) \\
-z^{\ast}(\tau ) & 0
\end{array} \right ) = - \frac{1}{2} Q(\tau ,u)\frac{\partial Q(\tau ,u)}
{\partial u} ,
\end{equation}
which when inserted into (\ref{simple}) yields the expression for
$S_B$ in one of its final forms
\begin{equation}
S_B = \int_0^{\beta} d\tau \int_0^1 du \left [\frac{S}{2} \mbox{Tr} \left (
Q (\tau ,u )
\frac{\partial Q (\tau ,u )}{\partial u} \frac{\partial Q (\tau ,u )}
{\partial \tau} \right ) \right ]
\label{string}
\end{equation}
An expression for $S_B$ solely in terms of $\bn ( \tau , u)$ can be
obtained by substituting in (\ref{adjoint}); this yields finally:
\begin{equation}
S_B = i S \int_0^{\beta} d\tau \int_0^1 du~ \bn \cdot \left(
\frac{\partial \bn}{\partial u} \times \frac{\partial \bn}{\partial \tau}
\right )
\label{stringn}
\end{equation}
This expression has a simple geometric interpretation: it is simply
$iS$ times
the area on the surface of a unit sphere
swept out by the string connecting $\bn ( \tau)$ to the
North Pole - this is also the oriented area contained within the closed
loop defined by $\bn ( \tau )$, $0 \leq \tau \leq \beta$, $\bn (0) =
\bn (\tau )$ on the sphere. Note that the $u$ dependence has
dropped out of this result which only depends on the values of $\bn (
\tau , u=1)$.

\section*{\normalsize\bf 3. NON-LINEAR SIGMA MODEL}
\refstepcounter{section}

We shall now obtain a long-wavelength, large-$S$
description of the dynamics of $d$-dimensional
antiferromagnets in the vicinity of a ground state with collinear
long-range N\'{e}el order. This description takes the form of an
imaginary-time functional integral over an action which includes a
non-linear sigma model field theory in $d+1$ dimensions; this result
was first noted by Haldane~\cite{theta,ian1,haleck}. The
advantage of this approach is that it is the physically most
transparent, and often the simplest,
way of describing many of the results. Furthermore,
although the derivation depends upon a semiclassical, large-$S$
limit, all of the results have also been obtained in alternative
large-$N$ methods~\cite{self1,self2,self3}
which are valid well into the quantum-disordered
phase. The chief disadvantages of the present approach are
that it connect simply describe the appearance of spin-Peierls order in the
quantum-disordered phase, and it does not have a simple
extension to a description of the phases with short-range or
long-range incommensurate spin correlations

Let us consider an antiferromagnet on a
$d$-dimensional hypercubic lattice with only a nearest-neighbor
exchange interaction $J$. In the semiclassical limit, the
spins will predominantly orient themselves in opposite directions on
the two sublattices. Let $\bN ( \br , \tau )$ be a continuum
field of unit length
which describes the local orientation of this N\'{e}el ordering -
$\bN$ varies slowly on the scale of a lattice spacing, but the
values of $\bN$ on well separated points can be considerably
different, leaving open the possibility of a quantum disordered phase
with no long-range spin order.
It will turn out to be necessary to also include a component of the
spins which is perpendicular to the local orientation of the N\'{e}el
order - this is described by the continuum field ${\bf L} ( \br ,
\tau )$.
We have therefore on the site $i$ of the lattice
\begin{equation}
\bn (i) \approx \varepsilon_i \bN ( \br_i )
\sqrt{1 - a^{2d} \bL^2 (\br_i )} + a^d \bL (\br_i ) ,
\label{nlsexp}
\end{equation}
where $\varepsilon_i$ equals $\pm 1$ on the two sublattices, $a$ is
the lattice spacing, and
\begin{equation}
\bN^2 = 1~~~~~~~\bN \cdot \bL = 0~~~~~~~~\bL^2 \ll a^{-2d}
\end{equation}
These relationship holds for all values of $\tau$ and $u$, and $\bN ,
\bL$ satisfy the same boundary conditions in $u$ as $\bn$.
We insert the decomposition for $\bn$ into $\cH (\bn (\tau ))$ and
expand the result in gradients and powers of $\bL$. This yields
\begin{eqnarray}
\cH &=&  \frac{1}{2}
\int d^d \br \  \left [
JS^2 a^{2-d} (\nabla_{\br} \bN )^2 + 2dJS^2 a^{d} \bL^2
\right ] \nonumber \\
&\equiv & \frac{1}{2}
\int d^d \br \  \left [
\rho_s (\nabla_{\br} \bN )^2 + \chi_{\perp} S^2 \bL^2
\right ]
\label{nlshamil}
\end{eqnarray}
the second equation defines the thermodynamic spin-wave stiffness, $\rho_s$,
and the transverse susceptibility $\chi_{\perp}$. If we had used a
different form for $\cH$ with modified short-range exchange
interactions, the continuum limit of $\cH$ would have been the same
but with new values of $\rho_s$ and $\chi_{\perp}$.

To complete the expression for the coherent state path-integral in
the the continuum limit we also need the expression for $S_B$ in
terms of $\bN , \bL$. We insert (\ref{nlsexp}) into the
(\ref{stringn}) and retain terms upto linear order in $\bL$: this
yields
\begin{equation}
S_B = S_B^{\prime} + iS \int d^d \br \int_0^{\beta} d \tau \int_0^1 du \left[
\bN \cdot \left( \frac{\partial \bN}{\partial u} \times
\frac{\partial \bL}{\partial \tau} \right) +
\bN \cdot \left( \frac{\partial \bL}{\partial u} \times
\frac{\partial \bN}{\partial \tau} \right) +
\bL \cdot \left( \frac{\partial \bN}{\partial u} \times
\frac{\partial \bN}{\partial \tau} \right)  \right]
\end{equation}
where
\begin{equation}
S_B^{\prime} = i S \sum_i \varepsilon_i
\int_0^{\beta} d\tau \int_0^1 du~ \bN_i \cdot \left(
\frac{\partial \bN_i}{\partial u} \times \frac{\partial \bN_i}{\partial \tau}
\right )
\end{equation}
The continuum limit of $S_{B}^{\prime}$ is not simple due to the
presence of the rapidly oscillating prefactor $\epsilon_i$. In two
spatial dimensions, a careful
examination by several investigators~\cite{hopf} showed that this term
is identically zero for all smooth spin configurations. However when
singular spin configurations are permitted, there can be a net Berry
phase~\cite{hedge}; this will studied in Section 6.C. We also note
in passing that in one-dimensional antiferromagnets~\cite{theta} it is
$S_B^{\prime}$ which gives rise to the topological $\theta$-term.

The remainder of $S_B$ can be simplified further. Using the fact that
the vectors $\bL$, $\partial\bN / \partial \tau$, $\partial \bN /
\partial u$ are all perpendicular to $\bN$ and hence lie in a plane
and have a vanishing triple product we find
\begin{equation}
S_B = S_B^{\prime} + iS \int d^d \br \int_0^{\beta} d \tau \int_0^1 du \left[
\frac{\partial}{\partial\tau} \left(\bN \cdot \left( \frac{\partial
\bN}{\partial u} \times \bL \right)\right)
+\frac{\partial}{\partial u} \left(\bN \cdot \left( \bL \times \frac{\partial
\bN}{\partial u} \right)\right)\right]
\end{equation}
The total $\tau$ derivative yields 0 as all fields are periodic in
$\tau$, while the total $u$ derivative yields a surface contribution
at $u=1$. This gives finally
\begin{equation}
S_B = S_B^{\prime} - iS \int d^d \br \int_0^{\beta} d \tau
\bL \cdot \left(\bN \times  \frac{\partial
\bN}{\partial \tau} \right)
\label{nlsberry}
\end{equation}
Putting together (\ref{nlshamil}) and (\ref{nlsberry}) we obtain the
following continuum limit path-integral for the antiferromagnet
\begin{eqnarray}
Z &=& \int {\cal D} \bN {\cal D} \bL \exp (S_n )\nonumber\\
S_n &=& S_B^{\prime} - \frac{1}{2}
\int_0^{\beta} d \tau \int d^d \br \  \left [
\rho_s (\nabla_{\br} \bN )^2 + \chi_{\perp} S^2 \bL^2
 + 2 i S \bL \cdot \left( \bN \times \frac{\partial \bN}{ \partial \tau}
\right)
 \right ]
\end{eqnarray}
The functional integral over $\bL$ is simply a gaussian and can
therefore be carried out
\begin{equation}
S_n = S_{B}^{\prime} + \frac{1}{2} \int_0^{\beta} d\tau \int d^d \br
\rho_s~
\left [
(\nabla_{\br} \bN )^2 + \frac{1}{c^2} (\partial_{\tau}
\bN )^2
\right ] ,
\label{nlaction}
\end{equation}
where the spin-wave velocity $c = \sqrt{\rho_s \chi_{\perp}}$.
This is the action for a $d+1$ dimensional non-linear sigma model
with a residual Berry phase term. As already noted, the Berry phase
terms only makes a non-zero contribution for topologically
non-trivial spin configurations.

\section*{\normalsize\bf 4. $O(M)$ NON-LINEAR SIGMA MODEL FOR LARGE
$M$}
\refstepcounter{section}
We will now study the properties of the $O(M)$ non-linear sigma model
field theory in a $1/M$ expansion; such an expansion was carried out
for the two-dimensional model by Polyakov~\cite{polyakovbook}. The
results in this section represent work carried out with Jinwu
Ye~\cite{jinwu,jinwu2}.

As we have seen in the previous section, the $O(3)$
non-linear sigma model, combined with some additional Berry phase
terms, describes the long-wavelength dynamics of the
$SU(2)$ antiferromagnet in the semiclassical limit. The Berry phases
will be completely ignored in this section - their consequences will
be considered later. Furthermore, the $O(M)$ model for $M>3$ is not
simply related to any quantum antiferromagnet. Our main motivation
for looking at this model is that, with proper interpretation,
it presents the simplest way at
arriving at some of the basic results and exploring crude features of
the phase diagram of the $SU(2)$ antiferromagnet. The omited Berry
phases do induce important differences between the $SU(2)$
antiferromagnet and the $O(3)$ non-linear sigma model - these will be
explored in subsequent subsections.

In this section, we will deal exclusively with the following
non-linear sigma model field
theory
\begin{equation}
Z = \int {\cal D} n_{a} \delta ( n_a^2 -1 )
\exp \left( - \frac{M}{2g} \int d^2 \br
\int_0^{c \beta} d \ttau \left[
(\nabla_{\br} n_a )^2 + (\partial_{\ttau}
n_a )^2
\right ] \right),
\label{nlsz}
\end{equation}
where the index $a$ runs from $1$ to $M$.
This action can be obtained from the semiclassical action for the
antiferromagnet (\ref{nlaction}) by omiting the Berry phase term
$S_B^{\prime}$, and introducing the rescaled time $\ttau = c \tau$.
We will henceforth omit the tilde on the $\tau$ and use units in
which $c=1$ - it is trivial to reinsert appropriate factors of $c$
in the final results. The coupling constant $g$ is given by
\begin{eqnarray}
g &=& \frac{Mc}{\rho_s} \nonumber \\
&=& \frac{6a}{S} ~\mbox{for the large $S$ $SU(2)$ antiferromagnet in
$d=2$}
\end{eqnarray}

The large $M$ analysis of $Z$ begins with the introduction of the
rescaled field
\begin{equation}
\tn_a = \sqrt{M} n_a
\end{equation}
and the imposition of the constraint by a Lagrange multiplier
$\lambda$. This transforms $Z$ into
\begin{equation}
Z = \int {\cal D} \tn_{a} {\cal D} \lambda
\exp \left( - \frac{1}{2g} \int d^2 \br
\int_0^{\beta} d \tau \left[
(\nabla_{\br} \tn_a )^2 + (\partial_{\tau}
\tn_a )^2 + i \lambda (\tn_a^2 - M)
\right ] \right),
\end{equation}
This action is quadratic in the $n_a$, which can therefore be
integrated out. This induces an effective action for the $\lambda$
field which has the useful feature of having all its $M$ dependence
in a prefactor. Therefore, for large $M$ the $\lambda$ functional
integral can be evaluated by an expansion about its saddle point.

Let us look more carefully at the solution at $M=\infty$ when the
saddle point action is the exact answer. At the saddle point we
parametrize
\begin{equation}
i \langle \lambda \rangle = m^2
\end{equation}
where $m$ is a function of $g, T,$ and an upper cutoff which will be
determined below.
The correlator of the $n_a$ field is the stagered dynamic spin
susceptibility and is given by
\begin{equation}
\chi ( \bk , \wn ) = \langle n_a ( \bk , \wn ) n_a ( -\bk , -\wn ) \rangle =
\frac{g}{\bk^2 + \wn^2 + m^2}
\end{equation}
where we have introduced the momentum $\bk$ and the Matsubara
frequency $\wn$. Neutron scattering experiments yield a direct
measurement of the imaginary part of $\chi$ ($\chi^{\prime\prime}$)
for real frequencies. At
$M=\infty$, analytic continuation of the above result yields
\begin{equation}
\chi^{\prime\prime} ( \bk , \omega ) = \frac{g\pi}{2 \sqrt{\bk^2 +
m^2}} \left( \delta ( \omega - \sqrt{\bk^2 + m^2} ) - \delta ( \omega
+ \sqrt{\bk^2 + m^2} ) \right)
\end{equation}
Another quantity often acessed in neutron scattering is the momentum
integrated local susceptibility, $\chi_L$, which is given by
\begin{equation}
\chi_L^{\prime\prime} ( \omega ) =
\int \frac{d^2 \bk}{4 \pi^2} \chi^{\prime\prime} ( \bk , \omega ) =
\frac{g}{4} \left( \theta ( \omega - m) - \theta (-\omega - m) \right)
\end{equation}
where $\theta (x)$ is the stop function which is non-zero only for
positive $x$.
An important shortcoming of the $M=\infty$ result is the absence of
any damping which is always present at finite $T$ - this only appears
at order $1/M$ which will be discussed later. The damping will
broaden the delta function peaks in $\chi ( \bk , \omega )$ and fill
in the gap region ($|\omega | < m$) in $\chi_L ( \omega )$ at all
finite temperatures and all values of $g$ - true delta functions and
gaps can however survive at $T=0$.
We can also examine the spin structure factor, $S(\bk)$ which
is the equal-time spin-spin correlation function in momentum space
\begin{eqnarray}
S( \bk ) &=& T \sum_{\wn} \langle n_a ( \bk , \wn ) n_a ( - \bk , -\wn
) \rangle \nonumber \\
&=& \frac{g}{2\sqrt{\bk^2 + m^2}} \mbox{coth} \left(
\frac{\sqrt{\bk^2 + m^2}}{2 T} \right)
\end{eqnarray}
The spin-correlation length, $\xi$, is therefore given by
\begin{eqnarray}
\xi^2 &\equiv& - \frac{1}{S(0)} \left.
\frac{\partial S}{\partial \bk^2} \right|_{\bk = 0} \nonumber \\
&=& \frac{1}{m^2} \left( \frac{1}{2} + \frac{m/T}{2~\mbox{sinh}(m/T)}
\right)
\label{xim}
\end{eqnarray}
The factor multiplying $1/m^2$ decreases monotonically from 1 to
$1/2$ as $m/T$ increases from 0. Thus $m$ is essentially the inverse
spin correlation length, apart from an innocuous numerical factor
between 1 and $1/\sqrt{2}$.

We now determine $m$.
The saddle-point equation is
simply the constraint $n_a^2 = 1$, or more explicitly
\begin{equation}
T \sum_{\wn} \int \frac{d^2 \bk}{4 \pi^2} \frac{ g}{\bk^2 + \wn^2 +
m^2} = 1
\end{equation}
where $T=1/\beta$ is the temperature (we are using units in which
$k_B = \hbar = 1$).
It is easy to see that this equation is divergent in the ultraviolet,
and it is therefore necessary to introduce an ultraviolet cut-off in
the momentum integration. The most convenient method is to use a
Pauli-Villars cut-off $\Lambda$. This transforms the constraint
equation into
\begin{equation}
T \sum_{\wn} \int_0^{\infty} \frac{d^2 \bk}{4 \pi^2} \left(
\frac{ g}{\bk^2 + \wn^2 +
m^2} - \frac{g}{\bk^2 + \wn^2 + \Lambda^2 } \right) = 1
\end{equation}
The momentum integration can be carried out and yields
\begin{equation}
\frac{gT}{4\pi} \sum_{\wn} \mbox{ln} \left( \frac{\wn^2 +
\Lambda^2}{\wn^2 + m^2} \right) = 1
\end{equation}
The frequency summation can be done exactly to give
\begin{equation}
\frac{gT}{2 \pi} \mbox{ln} \left(
\frac{\mbox{sinh}(\Lambda/2T)}{\mbox{sinh}(m/2T)} \right) = 1
\end{equation}
Finally, we can solve for the dependence of $m$ on $T, g,$ and
$\Lambda$
\begin{equation}
m = 2T \mbox{Arcsinh} \left( \exp\left(-\frac{2\pi}{gT}\right)
\mbox{sinh} \left(\frac{\Lambda}{2T} \right) \right)
\label{mval}
\end{equation}
A remarkable amount of information is contained in this innocuous
looking equation.
By examining the $T \rightarrow 0$ limit of this equation, it is
immediately apparent that the behavior of $m$ is quite different
depending upon whether $g$ is smaller, larger, or close to a critical
value $g_c$ given by
\begin{equation}
g_c = \frac{4 \pi}{\Lambda}
\end{equation}
We examine the three different phases separately.
\section*{\normalsize\bf Ordered Phase - $g < g_c$}
For $g< g_c$ we find for small $T$ that
\begin{equation}
m \sim \xi^{-1} \sim 2 T \exp \left( - \frac{2\pi}{T} \left(
\frac{1}{g} - \frac{1}{g_c} \right) \right)
\end{equation}
Thus the correlation length diverges exponentially as $T \rightarrow
0$ and long-range-N\'{e}el order appears at $T=0$. A renormalization
group analysis of the non-linear sigma model by Chakraborty {\em et.
al.\/}~\cite{sudip} yielded the same functional dependence of the
correlation length on $T$; this agreement gives us some confidence on
the utility of the present large $M$ expansion. We refer the reader
to Ref.~\cite{sudip} for further discussion on the dynamic properties
of the ordered phase - the semiclassical approach discussed there is
a little more convenient for analyzing this region.
\section*{\normalsize\bf Quantum Disordered Phase - $g > g_c$}
We now find for small $T$ that
\begin{equation}
m \sim \frac{1}{2\xi} \sim \frac{4 \pi ( g - g_c )}{g g_c}
\label{nu1}
\end{equation}
with exponentially small corrections at low $T$.
In this case there is a finite correlation length at $T=0$.
Furthermore the excitation spectrum has an energy gap $=c m$. The
low-lying excitations are the massive spin-1 (for $O(3)$) $n_a$ quanta -they
are thus triply degenerate. There will also be a spinless collective
mode with a gap represented by the fluctuations of $\lambda$ about
its saddle-point value.

As $g$ approaches $g_c$ we expect on general scaling grounds that
\begin{equation}
\xi \sim | g - g_c |^{-\nu}
\label{xinu}
\end{equation}
Comparing with (\ref{nu1}) we have $\nu =1$ at $M=\infty$. We have
considered the $1/M$ corrections to this result. The structure of the
perturbation theory is very similar to the $1/M$ expansion for
soft-spin Ginzburg Landau models as discussed in the book by
Ma~\cite{skma}; we will therefore omit the details. We find
\begin{equation}
\nu = 1 - \frac{32}{3\pi^2 M}
\end{equation}

\section*{\normalsize\bf Critical Region - $g \approx g_c$}
In this region both $\delta g = g - g_c$ and $T$ are small, and their
ratio can be arbitrary. Before examining the result (\ref{mval}) let
us see what can be predicted on general scaling grounds for general
$M$. At $g=g_c$ and $T=0$ we have a quantum field theory at its
critical point. Turning on a finite $T$ is equivalent to placing this
critical system in a box which is finite in the imaginary time
direction, while tuning $g$ away from $g_c$ at $T=0$ induces a finite
correlation length which scales like (\ref{xinu}). The behavior of
$\xi$ for both $T$ finite and $g \neq g_c$ is therefore predicted by
the principles of finite-size scaling
(after restoring factors of $\hbar , c,$ and
$k_b$)
\begin{equation}
\xi = \frac{\hbar c}{k_B T} X \left( a_1 \frac{\mbox{sgn}(g-g_c)
|g-g_c |^{\nu}}{T} \right)
\label{eqnxi}
\end{equation}
where $a_1$ is the only non-universal, cutoff dependent quantity and
$X (x)$ is a universal function. $X (x)$ is a smooth function of
$x$ except at $x=0$ where it is continuous but not differentiable.
It is not difficult to see that the $M=\infty$
Eqns.~(\ref{xim}) and (\ref{mval})
can be written in this form with $\nu = 1$, $a_1 = 2\pi/g_c^2$ and
$X (x)$ is defined by the following equations
\begin{eqnarray}
X (x) &=& \frac{1}{f(x)} \left( \frac{1}{2} + \frac{f(x)}{2~
\mbox{sinh}(f(x))} \right)^{1/2} \nonumber \\
f(x) &=& 2~ \mbox{ln} \left( \frac{e^x + \sqrt{4 + e^{2x}}}{2} \right)
\end{eqnarray}
A plot of the function $X (x)$ is shown in Fig.~\ref{xiplot}.
\begin{figure}
\vspace{4in}
\caption{\sl
The universal function $X (x)$ for the critical behavior of the
spin-correlation length at $M=\infty$. $X(0)$ is close to, but not
exactly, unity}
\label{xiplot}
\end{figure}

The \underline{\sl quantum-critical} region is defined by the
inequality $|\delta g |^{\nu} /T < 1$ (Fig~\ref{qcritf1}). In this region the
antiferromagnet notices that it is finite in the time direction at a
scale which is shorter than which it notices that the coupling $g$ is
not exactly $g_c$. Thus the critical spin fluctuations are quenched
at an energy scale which is determined completely by the temperature
and not any underlying antiferromagnetic exchange constant. In this
regime we may simply put $x=0$ in the formulae in the previous
paragraph.
\begin{figure}
\vspace{3in}
\caption{\sl
Phase diagram of ${\cal H}$ in two dimensions.
The magnetic LRO can be either spin-glass or N\'{e}el, and is present
only at $T=0$.  The boundaries of the quantum-critical region are $T
\sim |g-g_c|^{z\nu}$.  For non-random ${\cal H}$ which have
commensurate, collinear, N\'{e}el LRO for $g<g_c$, all of the
quantum-disordered region ($g>g_c$) has spin-Peierls order at
$T=0$-this order extends to part of the quantum disordered region at
finite $T$.}
\label{qcritf1}
\end{figure}

The full dynamic susceptibility also satisfies scaling functions in
the critical region. For simplicity we will only consider them in the
quantum critical region with $\delta g = 0$: the extension to finite
$\delta g$ is straightforward, at least at $M=\infty$. Application of
finite size-scaling to the dynamic susceptibility yields
\begin{equation}
\chi ( \bk , \omega ) = \frac{a_2}{T^{(2-\eta)}} \Phi \left( \frac{\hbar c
\bk}{k_B T} , \frac{\hbar \omega}{k_B T} \right)
\label{scalek}
\end{equation}
where $\Phi ( x, y )$ is a universal scaling function of both
arguments, $a_2$ is the only non-universal quantity, and $\eta$ is the
usual critical exponent determining the decay of spin correlations at
criticality.
A normalization condition is necessary to fix the scale of $\Phi$:
the most convenient is to use
\begin{equation}
\left. \frac{\partial \Phi^{-1}}{\partial x^2} \right|_{x=0,y=0} = 1
\end{equation}
The scaling form for the local susceptibility $\chi_L$
is a little more subtle. If we perform the momentum
integration of (\ref{scalek}) we notice immediately that the integral
is ultraviolet divergent because for large $\bk$, $\chi \sim \bk^{-2
+ \eta}$ and $\eta > 0$ (see below) for the present model. However
this divergence is present only in the real part of $\chi$. The
imaginary part of $\chi$ will only involve excitations only on-shell
and will therefore have rapidly vanishing spectral weight as $|\bk |$
becomes much larger than $\omega/c$, leading to a convergent momentum
space integral. Therefore we have the scaling form
\begin{equation}
\chi_L^{\prime\prime} ( \omega ) = \int \frac{d^2 \bk}{4 \pi^2}
\chi^{\prime\prime} ( \bk , \omega ) = a_3~|\omega|^{\eta} F \left(
\frac{\hbar \omega}{k_B T} \right)
\end{equation}
where $F$ is a universal scaling function and $a_3$ is a
non-universal constant. A remarkable feature of this result is that
the energy scale for all the excitations in the system is set solely
by the absolute temperature and is independent of any microscopic
energy scale.

The results for the quantum-critical scaling functions at $M=\infty$
can be easily determined from the results of this section. We find
$\eta = 0$,
\begin{equation}
\Phi (x,y) = \frac{1}{x^2 + \Theta^2 - (y + i \epsilon)^2}
\end{equation}
where $\epsilon$ is a positive infinitesimal, and
\begin{equation}
F ( y ) = \frac{1}{4}\left( \theta(y - \Theta) - \theta(-y -
\Theta)\right).
\end{equation}
The quantity $\Theta$ is a pure number given by
\begin{equation}
\Theta = f(0) = 2~ \mbox{ln} \left( \frac{1 + \sqrt{5}}{2} \right)
\end{equation}

We have computed these scaling functions to order $1/M$~\cite{jinwu}.
This is a
fairly non-trivial calculation and details will not be presented
here. We present the results below which required about 40 hours of
computation on a vectorized computer.
Our results for $\mbox{Im} \Phi$ and $F$ for $M=3$ are
summarized in Figs.~\ref{qcritf2} and~\ref{qcritf3}.
Note that $\mbox{Im} \Phi$ had a well
defined (at least for large $|\bk|$ )
spin-wave peak at a frequency close to $\omega = c |\bk |$.
The peak is broadened due to a universal damping arising from
spin-wave interactions.
\begin{figure}
\vspace{4in}
\caption{\sl
The imaginary part of the universal susceptibility in the
quantum-critical
region,
$\Phi$, as a function of $x = \hbar c q/(k_B T)$ and $y = \hbar \omega
/ (k_B T) $ for a non-random square lattice AFM which undergoes a $T=0$
transition from N\'{e}el LRO to a quantum-disordered
phase. The results have been
computed in a $1/M$ expansion to order $1/M$ and evaluated for $M=3$.
The two-loop diagrams were analytically continued to real frequencies
and
the integrals then evaluated numerically.
The shoulder on the peaks is due to a threshold towards three
spin-wave decay.}
\label{qcritf2}
\end{figure}
\begin{figure}
\vspace{4in}
\caption{\sl
The imaginary part of the universal local susceptibility, $F$, for the
same
model as in the previous figure.
We have $F(y) = y^{-\mu} \int d\vec{x} ~\mbox{Im} \Phi (
\vec{x},
y )$.
The oscillations at large $y$ are due to a finite step-size in the
momentum integrations.}
\label{qcritf3}
\end{figure}
Analytic forms for $\Phi$ can be obtained in various
regimes.
We have
\begin{equation}
\mbox{Re} \Phi^{-1} = C_Q^{-2} + x^2 + \ldots~~~~\mbox{$x,y$ small}
\end{equation}
The universal number $C_Q$, to order $1/M$, is:
\begin{equation}
C_Q^{-1} = \Theta \left( 1 + \frac{0.13}{N} \right)
\end{equation}
$\mbox{Im} \Phi$ has a singular behavior for $x,y$ small:
\begin{equation}
\mbox{Im} \Phi (x=0, y) \sim \frac{1}{M}
\exp\left (-\frac{3 \Theta^2}{2 |y|} \right)
\end{equation}
while
\begin{equation}
\mbox{Im} \Phi ( y < x) \sim \frac{1}{M}
y \exp\left(-\frac{3 \Theta^2 }{2|x|}\right)
\end{equation}
These peculiar singularities are probably artifacts of the large $M$
expansion and occur only in the region $x,y < 1/M$
- the naive expectation that $\mbox{Im} \Phi \sim y$ for
small $y$ is probably correct.
With either $x,y$ large, $\Phi$ has the form
\begin{equation}
\Phi = \frac{D_{Q}}{(x^2 - y^2)^{1-\eta/2}} + \ldots~~~~~;~~~~~
D_Q = 1 - \frac{0.3426}{N}.
\end{equation}
The exponent $\eta$ has the known~\cite{abe} expansion
\begin{equation}
\eta = \frac{8}{3 \pi^2
M} - \frac{512}{27\pi^4 M^2 } > 0
\end{equation}
The scaling function for the local susceptibility,
$F(y)$, has the limiting forms
\begin{equation}
F(y) = \mbox{sgn}(y)\frac{0.06}{N} |y|^{1-\eta} ~~~
\mbox{$y \ll 1$}~~~;~~~
F(y) = \mbox{sgn}(y) \frac{D_Q}{4} \frac{\sin ( \pi \eta /2 )}{\pi \eta /2}
{}~~~\mbox{$y \gg 1$}
\end{equation}
As $\eta$ is small, $F$ is almost linear at small $y$.
Note that the finite temperature $1/M$ fluctuations have filled in
the gap in the spectral functions at all frequencies. These scaling
functions represent the universal dissipative dynamics of critical
spin wave fluctuations.

\section*{\normalsize\bf 5. QUANTUM CRITICAL DYNAMICS OF 2D
ANTIFERROMAGNETS}
\refstepcounter{section}

The universal dynamic scaling forms discussed in the last section in
the context of unfrustrated antiferromagnets can
in fact be generalized to describe the quantum critical region of a
large number of
frustrated and random quantum antiferromagnets. Of course, explicit
calculations of the exponents and scaling functions is much more
difficult and only a few results are currently available. There has
however been some very recent progress~\cite{qsg,qlrm} which will not
be reviewed here.

Our motivation for considering quantum critical fluctuations in
random, frustrated antiferromagnets is provided by recent neutron
scattering experiments on lightly-doped $La_{2-x}\-Sr_{x}\-Cu\-O_4$ where
$0.02 < x < 0.07$. In this doping range it has been noted by the
experimentalists~\cite{hayden1,keimer} that the dynamic spin susceptibility
measured as a function of frequency, $\omega$, and temperature $T$
follows the following scaling form with reasonable accuracy
\begin{equation}
\int \frac{d^2 \bk}{4 \pi^2} \chi^{\prime\prime} ( \bk , \omega )
= I ( | \omega | ) F \left( \frac{\hbar\omega}{k_b T} \right)
\end{equation}
Experimental results for the functions $F$, $I$ from
Ref.~\cite{keimer2}
are shown in
Figs~\ref{figkeim1} and~\ref{figkeim2}.
We will argue below that the theory predicts that the prefactor
$I$ must be of the form
\begin{equation}
I \sim |\omega |^{\mu}
\end{equation}
where the exponent $\mu $ is expected to satisy $-1 < \mu < 0$. A fit
of this form by B. Keimer~\cite{bernie}
to the experimental result yielded $\mu =
-0.41 \pm 0.05$.
\begin{figure}
\vspace{4in}
\caption{\sl
Neutron scattering results for the scaling function $F$ from
Ref.~[15]}
\label{figkeim1}
\end{figure}
\begin{figure}
\vspace{4in}
\caption{\sl
Neutron scattering results for the normalization factor $I$ from
Ref.~[15]}
\label{figkeim2}
\end{figure}

A remarkable feature of the above result for $\chi$
is that the frequency scale for the spin fluctuations is set
completely by the absolute temperature. The underlying exchange
constants appear to modify only the prefactor of the scaling form.
We have recently proposed~\cite{jinwu} that this scaling of the
frequency scale with temperature is a rather general property of spin
fluctuations of two-dimensional antiferromagnets in their
quantum critical region. Let us discuss this in the context of the
general Hamiltonian ${\cal H}$ which we will use to model the
lightly-doped cuprates. In the lightly doped region the holes
are localized at low temperatures, indicating that a suitable form of
${\cal H}$ will be sufficient to describe the spins. A specific form
has in fact been proposed by Gooding and Mailhot~\cite{gooding} and yielded
reasonable results on the doping dependence of the zero temperature
spin correlation length.

We will find it necessary to distinguish between two
different types of magnetic LRO:
\newline
(${\bf A}$)
N\'{e}el LRO, in which case
\begin{equation}
{\bf m}_i \sim e^{i {\bf Q} \cdot
{\bf R}_i}
\end{equation}
with ${\bf Q}$ the N\'{e}el ordering wavevector,
and
\newline
(${\bf B}$) spin-glass LRO, in which case ${\bf m}_i$ can have an
arbitrary dependence on $i$, specific to the particular realization of
the randomness. The lower critical dimension of the Heisenberg
spin-glass~\cite{young_binder}
may be larger than 3 - in this case
the spin-glass LRO will not survive to any finite $T$, even in
the presence of a coupling between the layers. This, however, does
not preclude the existence of spin-glass LRO at $T=0$.

Consider now a $T=0$ phase transition between the magnetic LRO and
the quantum-disorder, induced by varying a coupling constant
$g$ which is dependent on the ratio's of the $J_{ij}$ in ${\cal H}$.
Let the transition occur at
a critical value $g=g_c$. If the transition is second-order then there
will be
a diverging scale
\begin{equation}
\xi_g \sim |g - g_c
|^{-\nu}
\end{equation}
which is the distance at which the antiferromagnet first notices that
its properties are not critical; at distances shorter than $\xi_g$ the
spin fluctuations are indistinguishable from those at criticality.
For N\'{e}el
LRO, $\xi_g$ is identical at $T=0$ to the spin correlation length
$\xi$ considered in Section 4. Therefore
$1/\xi$ is the width of the peak in the spin structure-factor at
the ordering wavevector ${\bf Q}$.
The meaning of $\xi_g$ is more subtle for the case of spin-glass LRO.
As the condensate ${\bf m}_i$ is a random function of $i$, there will
be no narrowing of the peak in the structure-factor. However it is
possible to define a diverging length scale associated with
correlations the
Edwards-Anderson order-parameter $q_{EA}$; $\xi_g$ will therefore be
related to the decay of
certain four-spin correlation
functions~\cite{young_binder}.

At finite temperature and near $g=g_c$ we can
define a second length scale $\xi_T $ associated with the distances at
which thermal effects become apparent. The dependence of $\xi_T$ on
$T$ can be deduced from finite-size scaling. At $g=g_c$ the
antiferromagnet is a $d+1$ dimensional critical system which is
described by a scale-invariant field theory. However, it is not
necessary in general for the theory to be Lorentz invariant, as
distances along the imaginary time direction can scale with a
non-trivial exponent, $z$, with respect to spatial distances. The
exponent $z$ is also referred to as the dynamic critical exponent.
As $1/T$ acts as a finite-size in the time direction we therefore
expect
\begin{equation}
\xi_T \sim \frac{1}{T^{(1/z)}}
\end{equation}
Now imagine that the system is in a region
where $\xi_T < \xi_g$ (we assume implicitly that
both these scales are much larger
than the lattice spacing)
which
defines the quantum-critical region - see Fig.~\ref{qcritf1}. The boundaries of
this region are therefore specified by
\begin{equation}
T \sim |g - g_c |^{z\nu}
\end{equation}
Under these circumstances the antiferromagnet will notice the finite
temperature {\em before} it notices the deviation from the purely
critical behavior at $g=g_c$. The spin-fluctuations will therefore be
controlled by the repulsive flow away from the critical fixed point,
cutoff by a finite scale which is set by the temperature. Both these
properties are universal features of the critical fixed point and the
dynamic spin susceptibility is therefore expected to obey a universal
scaling form. We will study the scaling form separately for the
transition from the two different types of magnetic long-range-order
distinguished above:

\section*{\normalsize\bf 5.A N\'{e}el order}

This section generalizes the results of Section 4 to include the
effects of disorder. We emphasize however that the disorder is not
strong enough to destroy the small-$g$ \neel ordered phase.

The first question which must addressed is whether an infinitesimal
amount of disorder is relevant at the fixed point describing the
transition in the pure system. We begin by assuming that disorder is
irrelevant. In this case the critical exponents will be unmodified. In
particular the exponent $\nu$ will preserve the value $\nu = 0.705 \pm
0.005$~\cite{nupure} of the pure system. However by a result of Chayes
{\em et.al.\/}~\cite{chayes} the
value of $\nu$ in any random system must satisfy
\begin{equation}
\nu \geq \frac{2}{d}
\end{equation}
where $d$ is the dimensionality of the space over which disorder is
uncorrelated. In our case, disorder is uncorrelated only along the two
spatial directions, which leads to $\nu \geq 1$. As this inequality is
violated by the above value of $\nu$, our hypothesis is inconsistent.
Therefore disorder is a relevant perturbation and must modify the
universality class of the transition.

The random fixed-point is quite difficult to access in general, but a
few of its properties can be delineated by general considerations.
Firstly, the space-time anisotropy induced by disorder implies that
the critical-theory need not be Lorentz invariant (unlike Section 4)
and therefore
\begin{equation}
z \neq 1
\end{equation}
The results in Section 4 for the dynamic scaling form for $\chi ( \bk
, \omega)$ can now be easily generalized.
At $T=0$ the static spin susceptibility, $\chi$,
will have a divergence at $g=g_c$ and wavevector $\bk = {\bf Q}$:
\begin{equation}
\chi(\bk = {\bf Q}, \omega = 0 ) \sim \frac{1}{|g-g_c|^{\gamma}}
\end{equation}
with
$\gamma = (2 - \eta) \nu$. At finite $\bk, \omega, T$
finite-size scaling yields the scaling form~\cite{jinwu}
\begin{equation}
\chi ( \bk , \omega ) = \frac{a_1}{T^{(2-\eta)/z}} \Phi \left(
\frac{a_2 |\bk -{\bf Q}|}{ T^{1/z}}, \frac{\hbar \omega}{k_B T} \right)
\label{scale_q}
\end{equation}
where $a_1 , a_2$ are non-universal constants, and $\Phi$ is a
universal, complex function of both arguments. The deviations from
quantum-criticality lead to an additional dependence of $\Phi$ on
$\xi_T / \xi$: this number is small in the QC
region and has been set to 0.
The scaling form for the local susceptibility can be obtained by
integrating (\ref{scale_q}) over $\bk$. The integral over $\bk$ will
be convergent in the ultraviolet provided $\eta < 0$ - available
results, which are discussed below, strongly indicate that all random
systems in fact have $\eta < 0$. We may therefore freely integrate
over $\bk$ and obtain
\begin{equation}
\chi_L^{\prime\prime} (\omega ) = a_3~ |\omega|^{\mu}~ F \left(
\frac{\hbar \omega}{k_B T} \right)
\label{scale_loc}
\end{equation}
with
\begin{equation}
\mu = \eta/z,
\end{equation}
\begin{equation}
F (y) = y^{-\mu} \int \frac{d^2 {\bf x}}{4 \pi^2}
{}~ \mbox{Im} \Phi ( \vec{x}, y )
\end{equation}
a universal function, and
$a_3$ a non-universal number. The real part
$\chi_L^{\prime}$ will also obey an identical scaling form as long as
$\eta < 0$.
The asymptotic forms for $F$ for small and large arguments can be
deduced from general considerations. For large $y$, or $\hbar \omega
\gg k_B T$, we expect $\chi$ to become independent of $T$; therefore
\begin{equation}
F(y) \sim \mbox{sgn} (y) ~~~~~~\mbox{for $|y| \gg 1$}
\label{limith}
\end{equation}
For small $\omega$, but $T$ finite,
we expect that $\chi_L^{\prime\prime} \sim \omega$ - no anomalous
scaling with $\omega$ is expected at finite temperature. Therefore
\begin{equation}
F(y) \sim |y|^{1-\mu} \mbox{sgn} (y)  ~~~~~~\mbox{for $|y| \ll 1$}
\label{limitl}
\end{equation}
We emphasize again that all non-universal
energy scales only appear in the prefactor $a_3$ and the frequency
scale in $F$ is determined solely by $T$.

Let us now examine two simple realizations of randomness in ${\cal H}$
for which exponents can be estimated from the literature.
\newline
{\em (a) Weak bond randomness}
\newline
Take the simplest pure model which has a transition from \neel LRO to
quantum disorder - we will argue in Section 6 that this is the square lattice
spin-1/2 antiferromagnet with first ($J_1$) and second ($J_2$)
neighbor interactions - the $J_1 - J_2$ model. The transition
is expected to be described by the non-linear sigma model of Section 4.
(As noted in Section 6.C, the quantum-disordered phase of this model
possesses spin-Peierls LRO - it has been argued however that the
coupling to the spin-Peierls order is dangerously
irrelevant~\cite{jinwu,daniel,ganpathy}
and does not modify two-spin correlation
functions in the quantum-critical region; the spin-Peierls order will
therefore be neglected here.)
Now add a small fluctuation in the $J_1$ bonds of the $J_1 - J_2$
model : {\em i.e.\/} $J_1 \rightarrow J_1 + \delta
J_1$ where $\delta J_1$ is random, with zero mean, and r.m.s. variance $\ll
J_1$. This last condition ensures that a N\'{e}el LRO to quantum
disorder transition will continue to occur as a function of $J_2 /
J_1$. Moreover, as the disorder is weak, the mapping to the non-linear
sigma model of Section 3 will continue to work, leading now to a
modified partition function (\ref{nlsz}) with the coupling $g$ a
random function of $\br$.
A soft-spin version of this action has been examined by Dorogovstev
and Boyanovsky and Cardy~\cite{dbc}. They considered a field theory
with a $M$ component order parameter which had random
couplings in $d=4-\epsilon-\epsilon_{\tau}$ space dimensions and
$\epsilon_{\tau}$ time dimensions. The theory was then examined in a double
expansion in $\epsilon, \epsilon_{\tau}$~\cite{dbc}. The expansion is
poorly-behaved, and for the case of interest here ($M=3$, $\epsilon=1$,
$\epsilon_{\tau} = 1$) the random fixed-point has the exponent
estimates
\begin{equation}
\eta = -0.17,~~~~
z=1.21,~~~~\nu = 0.64,~~~~\mu=-0.15.
\end{equation}
Note ({\em i\/}) $\mu,\eta$ are negative, unlike
the pure fixed point and ({\em ii\/}) $\nu$ is smaller than $2/d$,
violating the required bound~\cite{chayes}. This latter discrepancy is
not a cause for great concern as it is clear from the low-order
results that the series is badly behaved and that there are
large higher-order corrections.
\newline
{\em (b) Static holes on square lattice vertices}
\newline
Consider next a ${\cal H}$ on the square lattice with only $J_1$
couplings, but with a small concentration of static, spinless holes
on the vertices; this model will
display a N\'{e}el LRO to quantum-disorder transition at a critical
concentration of holes. In the coherent-state path-integral
formulation of the pure model, discussed in Section 3,
each spin contributes a Berry phase
which is almost completely canceled in the continuum limit between
the contributions of the two sublattices. The model
with holes will have large regions with unequal numbers of spins on
the two sublattices: such regions will contribute a Berry phase which
will almost certainly be relevant at long distances. Therefore the
field theory of Ref.~\cite{dbc} is \underline{not} expected to describe the
N\'{e}el LRO to quantum-disorder transition in this case.
It appears instead to be a model which possesses relevant spin
configurations whic have complex weights in Euclidean time, and not
amenable to a field-theoretic analysis by existing methods.
The only available results are those of
Wan
{\em et. al.}~\cite{wan} who performed a cluster expansion in the
concentration of spins {\em i.e.\/} they expanded about configurations
with the maximum number of holes. Their series analysis yielded the
exponents
\begin{equation}
\eta = -0.6,~~~~ z=1.7,~~~~
\nu = 0.8,~~~~ \mu=-0.35.
\end{equation}
Note again that $\mu,\eta < 0$, although the violation
of $\nu > 2/d$ suggests problems with the series extrapolations.
%Finally, we have also
%considered~\cite{longpap} the consequences of mobile holes
%in a non-random AFM. The
%spin-waves and holes were described by the
%Shraiman-Siggia~\cite{boris}  field theory.
%Integrating out the fermionic holes, led to a spin-wave self-energy
%$\Sigma_{\hat{n}} \sim a_1 |\vec{q}-\vec{Q}|^2 + a_2 \omega_n^2 + \ldots$,
%($a_1 , a_2$ constants) at $g=g_c$, $T=0$; non-analytic $|\omega_n|$ terms
%appear
%only with higher powers of $|\vec{q}-\vec{Q}|, \omega_n$ indicating
%that the
%N\'{e}el LRO to QD transition has the same leading
%critical behavior as that in the undoped, non-random $J_1-J_2$ model
%above. The exponents and scaling functions are identical, but the
%corrections to scaling are different.
%
\section*{\normalsize\bf 5.B Spin-glass order}

We now turn to the case where the degree of frustration is large
enough to destroy the N\'{e}el LRO state and induce first a state with
spin-glass LRO. The transition to quantum-disorder occurs subsequently from the
spin-glass phase. We will argue below that the lightly doped cuprates
are better described by this scenario.

Consider then the scaling of $\chi$ in the quantum-critical region of
the spin-glass LRO to quantum-disorder transition. As the condensate
${\bf m}_i$ is a random function of $i$, we do not expect any singular
behavior as a function of $\bk$.
In particular, the
scaling form (\ref{scale_q}) will \underline{not} be obeyed.
However,
the local susceptibility $\chi_L ( \omega_n )$ will be quite sensitive
to the transition. In particular, $\chi_L$ measures on-site
spin-correlation functions
\begin{eqnarray}
\chi_L ( \omega_n )
 &\equiv& \int_0^{1/(k_B T)} d
\tau e^{i \omega_n \tau} C(\tau), \nonumber \\
C(\tau) &=&
\overline{ \langle {\bf S}_i (0) \cdot {\bf
S}_i ( \tau ) \rangle }
\label{ctau}
\end{eqnarray}
(where the bar represents an average over all the sites $i$) which
acquire long-range order in time in the spin-glass phase. Thus
in the $T=0$ spin-glass phase~\cite{young_binder}
\begin{equation}
\lim_{\tau\rightarrow\infty} C(\tau ) =
\overline{ {\bf m}_i^2 } = q_{EA} > 0
\end{equation}
where $q_{EA}$ is the Edwards-Anderson order-parameter. At the phase
transition this order-parameter will vanish as
\begin{equation}
q_{EA} \sim (g_c - g)^{\beta}
\end{equation}
At $T=0$, the dynamic-scaling hypothesis then implies that near
$g=g_c$
\begin{equation}
C ( \tau ) = (g_c - g)^{\beta} h ( \tau |g - g_c |^{z\nu} )
\end{equation}
where $h$ is a universal scaling function which tends to a constant
for large argument. For $\tau \ll |g - g_c|^{-z\nu}$, $C(\tau)$ should
be independent of $g-g_c$; a standard argument then
implies that at criticality
\begin{equation}
C( \tau ) \sim \frac{1}{\tau^{\beta/(z\nu)}}
\end{equation}
Using (\ref{ctau}) it follows then that $\chi_L^{\prime\prime}$ obeys
a scaling form identical to (\ref{scale_loc})
with the exponent $\mu$ now given by
\begin{equation}
\mu = - 1 + \frac{\beta}{z\nu}
\label{muval}
\end{equation}
The function $F$ continues to satisfy the asymptotic forms in
(\ref{limith}) and (\ref{limitl}).
We emphasize that the above scaling forms are valid, despite the
failure of the $\bk$ dependent scaling (\ref{scale_q}). As everything
is expected to be non-singular as a function of $\bk$, we in fact
expect (\ref{scale_loc}) to hold independently at \underline{all $\bk$}.

Until recently~\cite{qsg,qlrm}, there were
essentially no theoretical results for $\mu$ for a
transition in this class. The identity (\ref{muval}) clearly shows
$\mu > -1$. It is also plausible that $\mu < 0$. This follows from
numerical studies of the
quantum-disordered phase
of random, spin-1/2
antiferromagnets~\cite{bhatt_lee} which found a divergent
zero-temperature local susceptibility in the quantum-disordered phase.
The susceptibility at the critical point is therefore also expected to
be divergent, implying $\mu < 0$.

Finally, we return to a discussion of the experimental situation.
An important property of the experiments is that the spin correlation
length becomes independent of temperature, over a significant range of
low temperatures. This is clearly incompatible with the presence of a
$T=0$ transition from \neel LRO to quantum-disorder which would have
$\xi \sim 1/T$ at $g=g_c$. This leaves the remaining possibility of a
spin-glass LRO to quantum-disorder transition.

Further support for
this scenario is provided by the work of Gooding and
Mailhot~\cite{gooding}: they considered a model of the doped cuprates
in which the dopant holes are localized on the oxygen sites and the
$Cu-O-Cu$ complex on that bond has been replaced by a single spin-1/2
doublet. This model leads naturally to strong frustration in the
ground state. A simulation of the classical version led a $T=0$
correlation length which was in good agreement with the experimental
results. Further, their results seem to indicate that the in the 2D
system, \neel LRO is destroyed and spin-glass LRO appears at any
non-zero concentration of dopants.

The main experimental measurements which then need to be understood
are the value of $\mu = -0.41\pm 0.05$ and the functional form of $F$.
The fact that $\mu < 0$ verifies experimentally the theoretical
prediction that disorder will destroy the pure fixed point which had
$\mu > 0$. More detailed theoretical studies of quantum spin glass are
clearly called for~\cite{qsg,qlrm}.

\section*{\normalsize\bf 6. QUANTUM DISORDERED PHASES OF NON-RANDOM 2D
ANTIFERROMAGNETS}

We now turn to an examination of the structure of the possible
quantum-disordered phases of ${\cal H}$ in the absence of randomness.
Our study of these phases has so far been limited to the large $M$,
$O(M)$ non-linear sigma model in Section 4. The structure of the
quantum-disordered phase found there was rather simple: the ground
state was a featureless spin-fluid with massive integer spin
excitations. No half-integer spin excitations were found. We shall
find upon careful examination here that this simple picture is almost
never valid~\cite{self4,self5}.
The only exception will be even-integer antiferromagnets
on the square lattice which have a two-sublattice
\neel ground state in the classical
limit~\cite{aklt}. We note however that the results on the quantum-critical
dynamic scaling functions found above are not modified by any of the
effects to be discussed in this section: the properties of the
quantum-disordered phase are controlled by a strong-coupling fixed
point while quantum criticality is determined by the critical
fixed point~\cite{jinwu}.

A careful examination of the quantum-disordered phase requires an
approach which is designed explicitly to be valid in a region well
separated from \neel LRO; the non-linear sigma model is essentially a
semiclassical expansion and can only approach the disordered phase
from the the \neel state. To this end, we introduce the Schwinger
boson description~\cite{assa}.
For the group $SU(2)$ the complete set of $(2S+1)$
states (\ref{szstates}) on site $i$ are represented as follows
\begin{equation}
|S , m \rangle \equiv \frac{1}{\sqrt{(S+m)! (S-m)!}}
(b_{i\uparrow}^{\dagger})^{S+m}
(b_{i\downarrow}^{\dagger})^{S-m} | 0 \rangle.
\end{equation}
We have introduced two flavors of bosons on each site,
created by the canonical operator
$b_{i\alpha}^{\dagger}$, with $\alpha = \uparrow, \downarrow$, and
$|0\rangle$ is the vacuum with no bosons. The total number of bosons, $n_b$
is the same for all the states; therefore
\begin{equation}
b_{i\alpha}^{\dagger}b_{i}^{\alpha} = n_b
\label{boseconst}
\end{equation}
with $n_b = 2S$ (we will henceforth assume an implied summation over
repeated upper and lower indices). It is not difficult to see that the above
representation of the states is completely equivalent to the following
operator identity between the spin and boson operators
\begin{equation}
\hat{S}_{ia} =  \frac{1}{2}
b_{i\alpha}^{\dagger} \sigma^{a\alpha}_{\beta} b_{i}^{\beta}
\end{equation}
where $a=x,y,z$ and the $\sigma^a$ are the usual $2\times 2$ Pauli
matrices.
The spin-states on two sites $i,j$ can combine to form a singlet in a unique
manner - the wavefunction of the singlet state is particularly simple
in the boson formulation:
\begin{equation}
\left( \varepsilon^{\alpha\beta} b_{i\alpha}^{\dagger}
b_{j\beta}^{\dagger} \right)^{2S} |0\rangle
\end{equation}
Finally we note that, using the constraint (\ref{boseconst}), the
following Fierz-type identity can be established
\begin{equation}
\left( \varepsilon^{\alpha \beta}
b_{i \alpha}^{\dagger} b_{j \beta}^{\dagger} \right)
\left( \varepsilon_{\gamma \delta}
b_{i}^{\gamma} b_{j}^{\delta} \right) =
 - 2 {\bf S}_i \cdot {\bf S}_j
+ n_b^2 /2 + \delta_{ij} n_b
\label{su2}
\end{equation}
where $\varepsilon$ is the totally antisymmetric $2\times2$ tensor
\begin{equation}
\varepsilon = \left( \begin{array}{cc}
0 & 1 \\
-1 & 0 \end{array} \right)
\end{equation}
This implies that ${\cal H}$ can be rewritten in the form (apart from
an additive constant)
\begin{equation}
{\cal H} = - \frac{1}{2} \sum_{<ij>} J_{ij} \left( \varepsilon^{\alpha \beta}
b_{i \alpha}^{\dagger} b_{j \beta}^{\dagger} \right)
\left( \varepsilon_{\gamma \delta}
b_{i}^{\gamma} b_{j}^{\delta} \right)
\label{hafkag}
\end{equation}
This form makes it clear that ${\cal H}$ counts the number of singlet
bonds.

We have so far defined a one-parameter ($n_b$) family of models ${\cal
H}$ for a fixed realization of the $J_{ij}$. Increasing $n_b$ makes
the system more classical and a large $n_b$ expansion is therefore not
suitable for studying the quantum-disordered phase. For this reason we
introduce a second parameter - the flavor index $\alpha$ on the bosons
is allowed to run from $1 \ldots 2N$ with $N$ an arbitrary integer.
This therefore allows the bosons to transform under $SU(2N)$
rotations.
However the $SU(2N)$ symmetry turns out to be too large. We want to
impose the additional restriction that the spins on a pair of sites be
able to combine to form a singlet state, thus generalizing the
valence-bond structure of $SU(2)$ - this valence-bond formation is
clearly a crucial feature determining the structure of the quantum
disordered phase. It is well-known that this is
impossible for $SU(2N)$ for $N>1$ - there is no generalization of the
second-rank, antisymmetric, invariant
tensor $\varepsilon$ to general $SU(2N)$.

The proper generalization turns out to be to the group
$Sp(N)$~\cite{self4}. This
group is defined by the set
of $2N\times2N$ \underline{unitary} matrices $U$
such that
\begin{equation}
U^{T} \cj U = \cj
\label{dsp2}
\end{equation}
where
\begin{equation}
\cj_{\alpha\beta} = \cj^{\alpha\beta} = \left( \begin{array}{cccccc}
  & 1 & & & & \\
-1 &  & & & & \\
 & &  & 1  & & \\
 & & -1 &  & & \\
 & &  & & \ddots & \\
 & & & & & \ddots
\end{array} \right)
\end{equation}
is the generalization of the $\varepsilon$ tensor to $N>1$. It is
clear that $Sp(N) \subset SU(2N)$ for $N>1$, while $Sp(1) \cong SU(2)$.
The $b_{i}^{\alpha}$ bosons transform as the fundamental
representation of $Sp(N)$; the ``spins'' on the lattice
therefore belong to the symmetric product of $n_b$
fundamentals, which is also an irreducible representation.
Valence bonds
\begin{equation}
\cj^{\alpha\beta} b_{i\alpha}^{\dagger}
b_{j\alpha}^{\dagger}
\end{equation}
can be formed between any two sites; this operator is a
singlet under $Sp(N)$ because of (\ref{dsp2}). The form (\ref{hafkag})
of ${\cal H}$ has a natural generalization to general $Sp(N)$:
\begin{equation}
{\cal H}= -\sum_{i>j} \frac{J_{ij}}{2N} \left(
\cj^{\alpha\beta}
b_{i\alpha}^{\dagger} b_{j,\beta}^{\dagger} \right)
\left( \cj_{\gamma\delta} b_{i}^{\gamma} b_{j}^{\delta}
\right)
\label{hex}
\end{equation}
where the indices $\alpha,\beta,\gamma,\delta$ now run over $1\ldots
2N$.
We recall also that the constraint (\ref{boseconst}) must be imposed
on every site of the lattice.

We now have a two-parameter ($n_b , N$) family of models ${\cal H}$
for a fixed realization of the $J_{ij}$. It is very instructive to
consider the phase diagram of ${\cal H}$ as a function of these two
parameters (Fig.~\ref{phasediag}).
\begin{figure}[t]
\begin{picture}(400,275)
\thicklines
\put(30,30){\vector(1,0){350}}
\put(30,30){\vector(0,1){250}}
\multiput(30,30)(18,18){11}{\line(1,1){15}}
\put(185,165){\vector(1,1){50}}
\put(165,185){\vector(1,1){50}}
\put(225,235){\shortstack{\large\bf BOSONIC \\ \large\bf LARGE - $N$}}
\put(210,80){\vector(1,0){70}}
\put(290,72){\shortstack{\large\bf QUANTUM \\ \large\bf DIMERS }}
\put(80,180){\vector(0,1){70}}
\put(45,260){\makebox{\large\bf SEMICLASSICAL}}
\put(205,5){\makebox{\LARGE $N$}}
\put(5,155){\makebox{\LARGE $n_b$}}
\put(210,130){\shortstack{\LARGE\sl QUANTUM \\ \LARGE\sl DISORDERED }}
\put(90,205){\shortstack{\LARGE\sl  MAGNETIC \\ \LARGE\sl LRO}}
\end{picture}
\caption{\sl Phase diagram of the 2D $Sp(N)$ antiferromagnet
${\cal H}$ as a function of the ``spin'' $n_b$}
\label{phasediag}
\end{figure}

The limit of large $n_b$, with $N$ fixed leads to the semi-classical
theory. For the special case of $SU(2)$ antiferromagnets with a
two-sublattice collinear \neel ground state, the semiclassical
fluctuations are described by the $O(3)$ non-linear sigma model of
Sections 3,4. For other
models~\cite{self1,ian1,johan,premi,halpsas,dombread,joli},
the structure of the non-linear sigma
models is rather more complicated and will not be considered here.

A second limit in which the problem simplifies is $N$
large at fixed $n_b$~\cite{brad,self1}.
It can be shown that in this limit the ground
state is quantum disordered. Further, the
low-energy dynamics of ${\cal H}$ is described by an effective quantum-dimer
model~\cite{dan,self1}, with
each dimer configuration representing a particular pairing
of the sites into valence-bonds.
This model is itself described by a non-trivial many-body
Hamiltonian and has so far only be studied for
the case of the unfrustrated
square lattice antiferromagnet~\cite{dimernum,self3,fradki,rodolfo}.
We will not discuss it any further here, but existing results will be briefly
noted later.

The most interesting solvable limit is obtained by fixing the ratio of
$n_b$ and $N$
\begin{equation}
\kappa = \frac{n_b}{N}
\end{equation}
and subsequently taking the limit of large $N$~\cite{assa}; this limit will be
studied in this section in considerable detail. The implementation of
${\cal H}$ in terms of bosonic operators also turns out to be
naturally suited for studying this limit. The parameter $\kappa$
is arbitrary; tuning $\kappa$ modifies the slope of the line in
Fig.~\ref{phasediag} along which the large $N$ limit is taken. From
the previous limits discussed above, one might expect that the ground
state of ${\cal H}$ has magnetic LRO for large $\kappa$ and is
quantum-disordered for small $\kappa$.
We will indeed find below that for any set of $J_{ij}$ there is a
critical value of $\kappa = \kappa_c$ which separates the
magnetically ordered and the quantum disordered phase.

A powerful feature of the bosonic large-$N$ limit noted is the
existence of a second-order phase transition at $N=\infty$. In the
vicinity of the phase transition, we expect the physics to be
controlled by long-wavelength, low-energy spin fluctuations; the
large-$N$ method offers an unbiased guide in identifying the proper
low-energy degress of freedom and determines the effective action
controlling them. Having obtained a long-wavelength continuum theory
near the transition, one might hope to analyze the continuum theory
independently of the large-$N$ approximation and obtain results that
are more generally valid.

We will discuss the structure of the $N=\infty$ mean-field theory ,
with $n_b = \kappa N$ in
Section 6.A. The long-wavelength effective actions will be derived and
analyzed in Section 6.B. Finally topological Berry phase effects will
be considered in Section 6.C.

\section*{\normalsize\bf 6.A Mean-field theory}
We begin by analyzing ${\cal H}$ at $N = \infty$
with $n_b = \kappa N$. As noted above, this limit is most conveniently
taken using the bosonic operators.
We may represent the partition function of ${\cal H}$ by
\begin{equation}
Z = \int {\cal D} Q {\cal D} b  {\cal D} \lambda
\exp \left( - \int_0^{\beta} {\cal L} d\tau \right) ,
\end{equation}
where
\begin{displaymath}
{\cal L} = \sum_{i} \left [
b_{i\alpha }^{\dagger}  \left( \frac{d}{d\tau} + i\lambda_i \right)
b_i^{\alpha} - i\lambda_i n_b \right ]~~~~~~~~~~~~~~~~~~~~~~
\end{displaymath}
\begin{equation}
{}~~~~~~~~~~~+  \sum_{<i,j>} \left [
N \frac{J_{ij} |Q_{i,j} |^2}{2}
- \frac{J_{ij} Q_{i,j}^{\ast}}{2} \cj_{\alpha\beta} b_i^{\alpha}
b_j^{\beta}
+ H.c. \right] .
\label{zfunct}
\end{equation}
Here the $\lambda_i$
fix the boson number of $n_b$ at each site;
$\tau$-dependence of all fields is implicit; $Q$ was introduced by
a Hubbard-Stratonovich decoupling of ${\cal H}$.
An important feature
of the lagrangian
${\cal L}$ is its $U(1)$ gauge invariance under which
\begin{eqnarray}
b_{i\alpha}^{\dagger} & \rightarrow & b_{i\alpha}^{\dagger} (i)
\exp \left( i\rho_i (\tau ) \right ) \nonumber \\
Q_{i,j} &\rightarrow & Q_{i,j} \exp \left( - i \rho_i (\tau ) - i
\rho_j (\tau ) \right) \nonumber \\
\lambda_i & \rightarrow & \lambda_i + \frac{\partial \rho_i}{\partial
\tau} (\tau) .
\label{gaugetrans}
\end{eqnarray}
The functional integral over ${\cal L}$
faithfully represents the partition function as
long as we fix a gauge, {\em e.g.} by the condition $d\lambda /d\tau = 0$
at all sites.

The $1/N$ expansion of the free energy can be obtained by integrating out
of ${\cal L}$ the $2N$-component $b$,$\bb$ fields to leave an effective action
for $Q$, $\lambda$ having co-efficient $N$ (since $n_b \propto N$).
Thus the $N \rightarrow \infty$
limit is given by minimizing the effective action with respect to
``mean-field'' values of $Q = \mq$, $\lambda = \blam$ (we are
ignoring here the possibility of magnetic LRO which requires an
additional condensate $x^{\alpha} = \langle b^{\alpha} \rangle$ - this
has been discussed elsewhere~\cite{self4,self5,kagome}).
This is in turn equivalent to solving
the mean-field Hamiltonian
\begin{equation}
{\cal H}_{MF} =  \sum_{<i,j>} \left(
N \frac{J_{ij} |\mq_{ij} |^2}{2}
- \frac{J_{ij} \mq_{i,j}^{\ast}}{2} \cj_{\alpha\beta} b_i^{\alpha}
b_j^{\beta}
+ H.c. \right) .
+ \sum_{i} \blam_i (
b_{i\alpha }^{\dagger} b_{i}^{\alpha} -  n_b )
\end{equation}
This Hamiltonian is quadratic in the boson operators and all its
eigenvalues can be determined by a Bogouibov transformation. This
leads in general to an expression of the form
\begin{equation}
{\cal H}_{MF} = E_{MF}[ \mq , \blam] + \sum_{\mu} \omega_{\mu} [\mq , \blam]
\gamma_{\mu\alpha}^{\dagger} \gamma_{\mu}^{\alpha}
\end{equation}
The index $\mu$ extends over $1\ldots$number of sites in the system,
$E_{MF}$ is the ground state energy and is a functional of $\mq$,
$\blam$, $\omega_{\mu}$ is the eigenspectrum of excitation energies
which is a also a function of $\mq$, $\blam$, and the
$\gamma_{\mu}^{\alpha}$ represent the bosonic eigenoperators. The
excitation spectrum thus consists of non-interacting spinor bosons.
The ground state is determined by minimizing $E_{MF}$ with respect to
the $\mq_{ij}$ subject to the constraints
\begin{equation}
\frac{\partial E_{MF}}{\partial \blam_i } = 0
\end{equation}
The saddle-point value of the $\mq$ satisfies
\begin{equation}
\mq_{ij} = \langle \cj_{\alpha\beta} b_{i}^{\alpha} b_{j}^{\beta}
\rangle
\end{equation}
Note that $\mq_{ij} = - \mq_{ji}$ indicating that $\mq_{ij}$ is a
directed field - an orientation has to be chosen on every link.

We will now consider the ground state configurations of the $\mq$,
$\blam$ fields and the nature of the bosonic eigenspectrum for a
variety of non-random antiferromagnets:

\section*{\normalsize\sl 6.A.1 $J_1 - J_2 - J_3$ model}

This is the square lattice antiferromagnet with first ($J_1$), second
($J_2$), and
third ($J_3$) neighbor interactions~\cite{johan,premi}.
We examined the values of the energy
$E_{MF}$ for $\mq_{ij}$ configurations which had a translational
symmetry with two sites per unit cell. For all parameter values
configurations with a single site per unit cell were always found to
be the global minima. We will therefore restrict our attention to
such configurations. The $\blam_i$ field is therefore independent of
$i$, while there are six independent values of $\mq_{ij}$:
\begin{eqnarray}
\mq_{i,i+\hx} &\equiv& Q_{1,x} \nonumber \\
\mq_{i,i+\hy} &\equiv& Q_{1,y} \nonumber \\
\mq_{i,i+\hy+\hx} &\equiv& Q_{1,y+x} \nonumber \\
\mq_{i,i+\hy-\hx} &\equiv& Q_{1,y-x} \nonumber \\
\mq_{i,i+2\hx} &\equiv& Q_{3,x} \nonumber \\
\mq_{i,i+2\hy} &\equiv& Q_{3,y}
\end{eqnarray}
For this choice, the bosonic eigenstates are also eigenstates of
momentum with momenta $\bk$ extending over the entire first Brillouin
zone. The bosonic eigenenergies are given by
\begin{displaymath}
\omega_{\bk} = \left( \lambda^2 - |A_{\bk}|^2 \right)^{1/2}
\end{displaymath}
\begin{displaymath}
A_{\bk} = J_1 \left( Q_{1,x}\sin k_x + Q_{1,y}\sin k_y
\right)
+J_2 \left(
Q_{2,y+x}\sin (k_y + k_x ) +Q_{2,y-x}\sin (k_y - k_x )
\right)
\end{displaymath}
\begin{equation}
{}~~~~~~~~~~~~+J_3 \left(
 Q_{3,x}\sin (2k_x ) + Q_{3,y}\sin (2 k_y ) \right)
\label{omegak}
\end{equation}

We have numerically examined the global minima of $E_{MF}$ as a
function of the three parameters $J_2 /J_1$, $J_3 / J_1$, and
$N/n_b$~\cite{self4,self5}.
The values of the $\mq_{ij}$ at any point in the phase
diagram can then be used to classify the distinct classes of states.
The results are summarized in Figs.~\ref{sp2f1}-\ref{sp2f4} which show various
sections of the three-dimensional phase diagram. All of the phases
are labeled by the wavevector at which the spin
structure factor has a maximum. This maximum is a delta function for
the phases with magnetic LRO, while it is simply a smooth function of
$\bk$ for the
quantum disordered phases (denoted by SRO in Figs~\ref{sp2f1}-\ref{sp2f4}). The
location of this maximum will simply be twice the wavevector at which
$\omega_{\bk}$ has a mimimum: this is because the structure factor
involves the product of two bosonic correlation functions, each of
which consists of a propagator with energy denominator
$\omega_{\bk}$.
\begin{figure}
\vspace{5in}
\caption{\sl
Ground states of the $J_1 - J_2 - J_3$ model for $J_3=0$
as a function of $J_2/J_1$
and $N/n_b$ ($n_b = 2S$ for $SU(2)$). Thick
(thin) lines denote first (second) order transitions at $N=\infty$.
Phases are identified by the wavevectors at which they have
magnetic long-range-order (LRO) or short-range-order (SRO).
The links with $Q_p\neq 0$ in each SRO phase are shown. The
large $N/n_b$, large $J_2 / J_1$ phase has the two sublattices
decoupled at $N=\infty$; each sublattice has \neel-type SRO.
Spin-Peierls order at finite $N$ for odd $n_b$ is
illustrated by the thick, thin and dotted lines. The $(\pi , \pi)$-SRO and the
``decoupled'' states have line-type spin-Peierls order for $n_b
=2 $(mod 4) and are valence-bond-solids for $n_b=0 $(mod 4).
The $(0,\pi )$-SRO state is a valence-bond-solid for all even
$n_b$.}
\label{sp2f1}
\end{figure}
\begin{figure}
\vspace{5in}
\caption{\sl
As in the previous figure but as a function of $J_2 / J_1$ and $J_3/ J_1$
for ({\it a\/}) $N/n_b = 1$ and ({\it b\/}) $N/n_b =5$. The
inset in
({\it a\/}) shows the region at the tip of the arrow
magnified by 20:
a direct first-order transition from $(\pi , \pi)$-LRO to
$(0 , \pi)$-LRO occurs up to $J_3/ J_1 = 0.005$.}
\label{sp2f2}
\end{figure}
\begin{figure}
\vspace{5in}
\caption{\sl
As in the previous figure but for $J_3 / J_1 = 0.35$}
\label{sp2f3}
\end{figure}
\begin{figure}
\vspace{5in}
\caption{\sl
As in the previous figure but for $J_3 / J_2 = 0.5$}
\label{sp2f4}
\end{figure}

Each of the phases described below has
magnetic LRO for large $n_b / N$ and is quantum disordered for small
$n_b /N$. The mean-field result for the structure of all of the quantum
disordered phases is also quite simple: they are featureless spin
fluids with free spin-1/2 bosonic excitations (``spinons'') with energy
dispersion $\omega_{\bk}$ which is gapped over the entire Brillouin
zone. Notice that this result is quite different from that of the
$O(M)$ non-linear sigma model of Sections 3,4 which found only
integer spin excitations - the difference will
be reconciled later. Some of the quantum disordered phases possess a
broken lattice rotation symmetry even at $N=\infty$ - these will be
described below.
The mininum energy spinons lie at a wavevector $\bk_0$ and
$\omega_{\bk_0}$ decreases as $n_b / N$. The onset of magnetic LRO
occurs at the value of $n_b /N$ at which the gap first vanishes:
$\omega_{\bk_0} = 0$. At still larger values of $n_b / N$, we get
macroscopic bose condensation of the $b$ quanta at the wavevector
$\bk_0$, leading to magnetic LRO at the wavevector $2 \bk_0$.

We now turn to a description of the various phases obtained. They can
be broadly classified into two types:

\subsubsection*{\underline{\sl Commensurate collinear phases}}
In these states the wavevector $\bk_0$ remains pinned at a
commensurate point in the Brillouin zone, which is independent of the
values of $J_2 / J_1$, $J_3 / J_1$ and $n_b /N$. In the LRO phase the
spin condensates on the sites are either parallel or anti-parallel to
each other, which we identify as collinear ordering. This implies
that the LRO phase remains invariant under rotations about about the
condensate axis and the rotation symmetry is not completely broken.

Three distinct realizations of such states were found
\subsubsection*{1. $ (\pi , \pi )$}
This is the usual two-sublattice \neel state of the unfrustrated
square lattice and its quantum-disordered partner.
These states have
\begin{equation}
Q_{1,x}=Q_{1,y}\neq 0,~
Q_{2,y+x}=Q_{2,y-x}=Q_{3,x}=Q_{3,y}=0
\label{pipi}
\end{equation}
 From (\ref{omegak}), the
minimum spinon excitation occurs at $\bk_0 = \pm (\pi /2 , \pi /2)$.
The SRO states have no broken symmetry at $N=\infty$.
The boundary between the LRO and SRO phases occurs at
$N/n_b < 2.5$,
independent of $J_2 / J_1$ (Fig~\ref{sp2f1}). This last feature is surely
an artifact of the large $N$ limit.
Finite $N$ fluctuations should be stronger as
$J_2 / J_1$ increases, causing the boundary to bend a little
downwards to the
right.

\subsubsection*{2. $(\pi , 0)
$ or $ (0, \pi )$}
The $(0,\pi)$ states have
\begin{equation}
Q_{1,x}=0, Q_{1,y} \neq 0,
Q_{2,y+x} = Q_{2,y-x} \neq 0,\mbox{~and~} Q_{3,x} = Q_{3,y } =0
\end{equation}
and minimum energy spinons at $\bk_0 = \pm ( 0, \pi /2)$.
The degenerate
$(\pi , 0)$ state is obtained with the mapping $x
\leftrightarrow y$. The SRO state has a two-field degeneracy due to
the broken $x \leftrightarrow y$ lattice symmetry.
The LRO state again has two-sublattice collinear \neel order, but the
assignment of the sublattices is different from the $(\pi , \pi )$
state. The spins are parallel along the $x$-axis, but anti-parallel
along the $y$-axis.

An interesting feature of the LRO state here is the occurence of
\underline{order-from-disorder}~\cite{henley}. Recall that the classical limit
($n_b / N = \infty$) of
this model has an accidental degeneracy for
$J_2 / J_1 > 1/2$: the ground state
has independent collinear N\'{e}el order on each of the $A$ and $B$
sublattices, with the energy independent of the angle between the
spins on the two sublattices. Quantum fluctuations are included
self-consistently in the $N=\infty$, $n_b / N$ finite, mean-field
theory, and lead to an alignment of the spins on the sublattices and
LRO at $(0 , \pi)$. The orientation of the ground state has thus been
selected by the quantum fluctuations.

The $(0 , \pi)$ states are
separated from the $(\pi , \pi)$ states by a first-order transition.
In particular, the spin stiffnesses of both states remain finite at
the boundary between them. This should be distinguished from the
classical limit in which the stiffness of both states vanish at their
boundary $J_2 = J_1 /2$; the finite spin stiffnesses are thus
another manifestation of order-from-disorder. At a point well away
from the singular point $J_2 = J_1 /2$, $n_b / N = \infty$ in Fig~\ref{sp2f1},
the stiffness of both states is of order $N ( n_b / N)^2$ for
$N=\infty$ and large $n_b / N$; near this singular point however the
stiffness is of order $N ( n_b / N)$ is induced purely by quantum
fluctuations. These results have since also been obtained by a
careful resummation of the semiclassical expansion~\cite{chubukov,mila}.

\subsubsection*{3. ``Decoupled''}
For $J_2 / J_1 $  and $N/n_b$ both
large, we have a ``decoupled'' state (Fig~\ref{sp2f1})
with
\begin{equation}
Q_{2,y+x} = Q_{2,y-x} \neq 0~\mbox{and}~
Q_1=Q_3=0.
\end{equation}
In this case
$Q_p$ is non-zero only between sites on the same
sublattice. The two sublattices have \neel type SRO
which will be coupled by finite $N$ fluctuations.
The $N=\infty$
state does not break any lattice symmetry.
This state has no LRO partner.

\subsubsection*{\underline{\sl Incommensurate phases}}
In these phases the wavevector $\bk_0$ and the location of the
maximum in the structure factor move continuously with the
parameters~\cite{shraisiggkane}.
The spin-condensate rotates with a period which is not
commensurate with the underlying lattice spacing. Further the spin
condensate is {\em coplanar\/}: the spins rotate within a given plane
in spin space and are not collinear. There is this no spin rotation axis
about which the LRO state remains invariant.

Further, no states in which the spin condensate was fully three
dimensional (``double-spiral'' or chiral states) were found; these
would be associated with complex values of $Q_p$. All the saddle
points possesed a gauge in which all the $Q_p$ were real.
Time-reversal symmetry was therefore always preserved in all the SRO
phases of Figs~\ref{sp2f1}-\ref{sp2f4}.

The incommensurate phases occur only in models with a finite $J_3$
(Figs~\ref{sp2f2}-\ref{sp2f4}). There were two realizations:
\subsubsection*{1. $(\pi , q)$ or $(q, \pi)$}
Here $q$ denotes a wavevector which varies continuously between $0$
and $\pi$ as the parameters are changed. The $(q , \pi )$ state has
\begin{equation}
Q_{1,x}\neq Q_{1,y} \neq 0,~
Q_{2,x+y} = Q_{2,y-x} \neq 0,~Q_{3,x} \neq 0~\mbox{and}~
Q_{3,y}=0;
\end{equation}
the
degenerate $(\pi,q)$ helix is obtained by the mapping $x
\leftrightarrow
y$. The SRO state has a two-fold degeneracy due to the broken $x
\leftrightarrow y$ lattice symmetry.
\subsubsection*{2. $(q, q)$ or $(q, -q)$}
The $(q,q)$ state has
\begin{equation}
Q_{1,x}=Q_{1,y} \neq 0,~
Q_{2,x+y}\neq 0,~Q_{2,y-x} = 0,~Q_{3,x}=Q_{3,y}\neq 0;
\end{equation}
this
is degenerate with the $(q, -q)$ phase and SRO state therefore has a
two-fold degeneracy due to a broken lattice reflection symmetry.

Note that the broken discrete symmetries in states
with SRO at $(0 , \pi )$ and $(q , \pi)$ are identical:
both are two-fold degenerate due to a breaking of the
$x \leftrightarrow y$ symmetry. The states
are only distinguished by a non-zero value of $Q_3$
in
the $(q, \pi )$ phase and the accompanying incommensurate
correlations in the spin-spin correlation functions.
However $Q_3$ is gauge-dependent
and so somewhat
unphysical as an order parameter.
In the absence of any further fluctuation-driven lattice
symmetry
breaking (see below), the transition between SRO at $(0, \pi
)$
and $(q , \pi)$ is an example of a {\em disorder
line}~\cite{disorder}; these are lines at which
incommensurate correlations first turn on.

An interesting feature of Figs~\ref{sp2f3}-\ref{sp2f4} is that the
commensurate
states squeeze out the incommensurate phases as $N/n_b$
increases in both phase diagrams.
We expect that this suppression of incommensurate order by
quantum
fluctuations is
a general feature of frustrated
antiferromagnets. This result is also consistent with the natural
hypothesis that the states of the large $N$, fixed, but small $n_b /
N$ should be consistent with the states of the large $N$, fixed $n_b$
theory (Fig.~\ref{phasediag}). This latter limit is described by the
quantum-dimer model~\cite{dan} which is necessarily associated only with
commensurate states.

\section*{\normalsize\sl 6.A.2 Triangular and Kagom\'{e} Lattices}
We have also examined the mean-field equations for the
nearest-neighbor antiferromagnet on the triangular~\cite{yoshi} and
kagom\'{e} lattices in considerable detail~\cite{kagome}.
In both cases we found that the SRO phases had
the {\em full symmetry of the underlying lattice}. They differ in this manner
from all of the SRO phases of the square lattice. Further the
$\mq_{ij}$ were real on every link and had the same magnitude
$|\mq_{ij} | = Q$. The only
remaining degree of freedom is that associated with assigning an
orientation to each link: minimization of the energy determined a
unique orientation upto gauge-equivalent configurations.

On the triangular lattice the spectrum of the free spin-1/2 bosonic
spinon excitations was found to be given by
\begin{equation}
\omega ({\bf k} ) = \left( \lambda^2 - J^2 Q^2 ( \sin k_1  + \sin k_2
+ \sin k_3  )^2 \right)^{1/2}
\end{equation}
where the momentum ${\bf k}$ ranges over the first Brillouin zone of the
triangular lattice and
\begin{equation}
k_p = {\bf k} \cdot \hat{e}_p
\label{defki}
\end{equation}
with the $\hat{e}_p$ being unit vectors of length $a$ pointing along
the directions of the bonds on the triangular lattice
\begin{eqnarray}
\hat{e}_1 &=& a(1/2 , \sqrt{3}/2 ) \nonumber \\
\hat{e}_2 &=& a(1/2 , -\sqrt{3}/2 ) \nonumber \\
\hat{e}_3 &=& a(-1 , 0 )
\label{defei}
\end{eqnarray}
A surface plot of this spectrum is shown in Fig~\ref{trianglespin}
for $n_b /N = 0.25$.
\begin{figure}
\vspace{4in}
\caption{\sl
 Momentum dependence of the energy, $\omega ( {\bf k} )$ of the
lowest excited spinon state for the quantum-disordered ground state
of the triangular lattice quantum antiferromagnet at
$\kappa = 0.25$.
The minimum excitations
are the spinons at ${\bf k} = \tilde{{\bf k}}_1 = (4\pi /3a) (1, 0)$
and ${\bf k} =
\tilde{{\bf k}}_2 =
(4\pi /3a ) (-1 , 0)$ and other points separated from $\tilde{{\bf
k}}_1 ,
\tilde{{\bf k}}_2$
by vectors of the reciprocal lattice}
\label{trianglespin}
\end{figure}
The minimum-energy
spinons are those at ${\bf k}_{01} = (4\pi /3a) (1, 0)$ and
${\bf k}_{02} =
(4\pi /3a ) (-1 , 0)$ and other points separated from these points
by vectors of the reciprocal lattice generated
by ${\bf G}_1 =
(4\pi / \sqrt{3} a) ( 0, 1)$ and ${\bf G}_2 =
(4\pi / \sqrt{3} a) ( \sqrt{3}/2 , -1/2)$.
For $n_b /N > 0.34$ this state acquires magnetic LRO with the
conventional three-sublattice \neel ordering of the triangular
lattice. The spins are {\em coplanar} pointing towards the vertices of
an equilateral triangle.

Very similar results were obtained on the kagom\'{e} lattice. As this
is not a Bravais lattice, it is necessary to introduce three sites per
unit cell although no symmetry of the lattice is broken by the SRO
state.
To determine the free spin-1/2 bosonic excitation spectrum we need the
following matrix
\begin{equation}
P ( {\bf k} ) = -iJQ \left( \begin{array}{ccc}
0 & \sin k_1 & \sin k_3 \\
\sin k_1 & 0 & \sin k_2 \\
\sin k_3 & \sin k_2 & 0
\end{array} \right).
\label{pmatdef}
\end{equation}
(the $k_p$ were defined in Eqs.~(\ref{defki}) and (\ref{defei})) and
to solve the eigenvalue equation
\begin{equation}
P^{\dagger} ( {\bf k} ) P ({\bf k} ) \varphi_{\mu} ( {\bf k} ) =
p_{\mu}^2 ( {\bf k} ) \varphi_{\mu} ( {\bf k} )
\label{peigdef}
\end{equation}
where $p_{\mu}^2$, $\mu = 1,2,3$ are the eigenvalues and
$\varphi_{\mu} ( {\bf k} )$ the eigenvectors. Finally the bosonic
eigenspectrum is given by
\begin{equation}
\omega_{\mu} ( {\bf k} ) = \left( \lambda^2 - p_{\mu}^2 ( {\bf k} )
\right)^{1/2}
\label{ompdef}
\end{equation}
The lowest energy spinon spectrum for $n_b / N = 0.35$ is plotted in
Fig~\ref{kagomespin}.
\begin{figure}
\vspace{4in}
\caption{\sl
Momentum dependence of the energy, $\omega ( {\bf k} )$ of the
lowest excited spinon state for the quantum-disordered ground state
(which has $Q_1=-Q_2$) of the kagom\'{e} lattice quantum antiferromagnet at
$\kappa = 0.35$. The
energy is measures in units of $J/2$, and $a$ is the nearest-neighbor
spacing on the kagom\'{e} lattice.
The minimum excitations
are the spinons at ${\bf k} = {\bf k}_1 = (2\pi /3a) (1, 0)$ and ${\bf k} =
{\bf k}_2 =
(2\pi /3a ) (-1 , 0)$ and other points separated from ${\bf k}_1 , {\bf k}_2$
by vectors of the reciprocal lattice}
\label{kagomespin}
\end{figure}
The minimum energy
spinons are
at ${\bf k} = {\bf k}_1 = (2\pi /3a) (1, 0)$ and
${\bf k} = {\bf k}_2 =
(2\pi /3a ) (-1 , 0)$ and other points separated from ${\bf k}_1 , {\bf k}_2$
by vectors of the reciprocal lattice generated by ${\bf G}_1 =
(2\pi / \sqrt{3} a) ( 0, 1)$ and ${\bf G}_2 =
(2\pi / \sqrt{3} a) ( \sqrt{3}/2 , -1/2)$.
There is no quite good evidence that the spin-1/2, nearest-neighbor,
kagom\'{e} antiferromagnet is quantum
disordered~\cite{zengelser,rajivkag}. The present
calculation should therefore be relevant for this system.

The spinon gap vanishes and magnetic LRO appears for $n_b / N > 0.53$.
The magnetic LRO is the configuration identified as the coplanar
``$\sqrt{3} \times \sqrt{3}$'' structure in the
literature~\cite{chubkag,huserut,kallin,shender}. The huge
accidental degeneracy of the classical ground state on the kagom\'{e}
lattice is completely lifted by the quantum fluctuations, and unique
(upto global spin rotations) magnetic structure is obtained as the
ground state. Note the rather natural way in which this happened
directly in the the mean-field theory. The present large-$N$ approach
thus seems to be ideally suited to examining order-from-disorder
issues in frustrated antiferromagnets.

\section*{\normalsize\bf 6.B Fluctuations - long wavelength effective
actions}
We now extend the analysis of Section 6.B beyond the mean-field theory
and examine the consequences of corrections at finite $N$. The main
question we hope to address are:
\begin{itemize}
\item
The mean-field theory yielded an excitation spectrum consisting of
free spin-1/2 bo\-son\-ic spinons. We now want to understand the nature of
the forces between these spinons and whether they can lead to
confinement of half-integer spin excitations.
\item
Are there any collective excitations and does their dynamics modify in
any way the nature of the ground state ?
\end{itemize}

The structure of the fluctuations will clearly be determined by the
low-energy excitations about the mean-field state. We have already
identified one set of such excitations: spinons at momenta near
mimima in their dispersion spectrum, close to the onset of the magnetic
LRO phase whence the spinon gap vanishes. An additional set of
low-lying spinless excitations can arise from the fluctuations of the
$Q_{ij}$ and $\lambda_i$ fields about their mean-field values. The
gauge-invariance (\ref{gaugetrans}) will act as a powerful restriction
on the allowed in the effective action for these spinless fields. We
anticipate that the only such low-lying excitations are associated
with the $\lambda_i$ and the {\em phases\/} of the $Q_{ij}$. We
therefore
parametrize
\begin{equation}
Q_{i,i+\hat{e}_p} = \bar{Q}_{i,i+\hat{e}_p}
\exp\left( -i \Theta_p \right)
\label{qtheta}
\end{equation}
where the vector $\hat{e}_p$ connects the two sites of the lattice
under consideration,
$\bar{Q}$ is the mean-field value, and $\Theta_p$ is a real phase.
The gauge invariance (\ref{gaugetrans} implies that the
effective action for the $\Theta_p$ must be invariant under
\begin{equation}
\Theta_p \rightarrow \Theta_p + \rho_i + \rho_{i+\hat{e}_p}.
\end{equation}
Upon performing a Fourier transform, with the link variables $\Theta_p$
placed on the center of the links, the gauge invariance takes the
form
\begin{equation}
\Theta_p ( {\bf k} ) \rightarrow
\Theta_p ( {\bf k} ) + 2 \rho ({\bf k} ) \cos( k_p /2)
\label{thetatr}
\end{equation}
where $k_p = \bk \cdot \hat{e}_p$.
This invariance implies that the effective action
for the $\Theta_p$, after integrating out the $b$ quanta,
can only be a function of the following gauge-invariant
combinations:
\begin{equation}
I_{pq} = 2 \cos(k_q /2) \Theta_p ({\bf k} ) -
2 \cos (k_p /2) \Theta_q ({\bf k} )
\label{ipqdef}
\end{equation}
We now wish to take the continuum limit at points in the Brillouin zone
where the action involves only gradients of the $\Theta_p$ fields and
thus has the possibility of gapless excitations. This involves
expanding about points in the Brillouin zone where
\begin{equation}
\cos ( k_p / 2) = 0~\mbox{for the largest numbers of $\hat{e}_p$}
\label{lowenergy}
\end{equation}
We will apply this general principle to the models
considered in Sec. 6.A.

\section*{\normalsize\sl 6.B.1 $J_1 - J_2 - J_3$ Model}

We begin by examining the $(\pi , \pi)$-SRO phase.
As noted in (\ref{pipi}), this phase has the mean field
values
$Q_{1,x} = Q_{1,y} \neq 0$, and all other $\mq_{ij}$ zero.
Thus we need only examine the condition (\ref{lowenergy}) with
$\hat{e}_p = \hat{e}_x , \hat{e}_y$. This uniquely identifies the
point $\bk = \bG = (\pi , \pi)$ in the Brillouin zone. We therefore
parametrize
\begin{equation}
\Theta_x ( \br ) = i e^{i \bG \cdot \br} A_x ( \br )
\label{thetax}
\end{equation}
and similarly for $\Theta_y$; it can be verified that both $\Theta$
and $A_x$ are real in the above equation.
We will also be examining invariances of
the theory under gauge transformations near $\bG$: so we write
\begin{equation}
\rho ( \br ) = e^{i \bG \cdot \br } \varphi ( \br )
\label{rhophi}
\end{equation}
It is now straightforward to verify that the gauge transformations
(\ref{thetatr}) are
equivalent to
\begin{equation}
A_x \rightarrow A_x + \partial_x \varphi
\end{equation}
and similarly for $A_{y}$. We will also need in the continuum limit
the component of $\lambda$ near the wavevector $\bG$. We therefore
write
\begin{equation}
\lambda_i = \bar{\lambda} + i e^{i \bG \cdot \br}  A_{\tau} ( \br_i )
\label{anslam}
\end{equation}
Under gauge transformations we have
\begin{equation}
A_{\tau} \rightarrow A_{\tau} + \partial_{\tau} \varphi
\end{equation}
Thus $A_x$, $A_y$, $A_{\tau}$ transform as components of a continuum
$U(1)$ vector gauge field.

We will also need the properties of the boson operators under
$\varphi$. From (\ref{gaugetrans}) and (\ref{rhophi}) we see that the
bosons on the two sublattices ($A,B$) with opposite charges $\pm 1$:
\begin{eqnarray}
b_{A} &\rightarrow& b_{A} e^{i\varphi} \nonumber \\
b_{B} &\rightarrow& b_{B} e^{-i\varphi}
\end{eqnarray}
Finally, we note that the bosonic eigenspectrum has a minimum near $\bk
= \bk_0 = (\pi/2 , \pi/2 )$; we therefore parametrize
\begin{eqnarray}
b_{Ai}^{\alpha} &=& \psi_1^{\alpha} (\br_i ) e^{i \bk_0 \cdot \br_i }
\nonumber \\
b_{Bi}^{\alpha} &=& -i \cj^{\alpha\beta} \psi_{1\beta} (\br_i )
e^{i \bk_0 \cdot \br_i }
\label{bosepar}
\end{eqnarray}

We insert the continuum parametrizations (\ref{thetax}),
(\ref{anslam})
and (\ref{bosepar}) into the functional integral (\ref{zfunct}),
perform a gradient expansion, and transform the Lagrangian ${\cal L}$ into
\begin{displaymath}
{\cal L} = \int \frac{d^2 r}{a^2} \left [
\psi_{1\alpha}^{\ast} \left( \frac{d}{d\tau} + i A_{\tau}
\right)
\psi_1^{\alpha} +
\psi_{2}^{\alpha\ast} \left( \frac{d}{d\tau} - i A_{\tau}
\right)
\psi_{2\alpha} + \bar{\lambda} \left( |\psi_1^{\alpha} |^2
+ |\psi_{2\alpha} |^2 \right) \right.
\end{displaymath}
\begin{displaymath}
{}~~~~~~~~~~~~~~~~~~~~~~
-4 J_1 \bqo \left ( \psi_1^{\alpha}\psi_{2\alpha} +
\psi_{1\alpha}^{\ast}\psi_2^{\alpha\ast}
\right )
+  J_1 \bqo a^2 \left [
\left ( \vec{\nabla}  + i  \vec{A} \right ) \psi_1^{\alpha}
\left ( \vec{\nabla}  - i \vec{A} \right ) \psi_{2\alpha} \right.
\end{displaymath}
\begin{equation}
{}~~~~~~~~~~~~~~~~~~~~~~~~~~~~~~~~~~~~~~~~~~~~~~~~~~~~~~~~~~~~~
+ \left. \left ( \vec{\nabla} - i \vec{A} \right )
\psi_{1\alpha}^{\ast}
\left ( \vec{\nabla} + i \vec{A} \right )
\psi_2^{\alpha\ast} \right ]  \Biggr]
\label{charge}
\end{equation}
We now introduce the fields
\begin{eqnarray*}
z^{\alpha} & = & (\psi_1^{\alpha} +
\psi_2^{\alpha\ast})/\sqrt{2} \\
\pi^{\alpha} & = & (\psi_1^{\alpha} -
\psi_2^{\alpha\ast})/\sqrt{2} .
\end{eqnarray*}
 From Eqn (\ref{charge}), it is clear that the
the $\pi$ fields turn out to have mass $\blam + 4 J_1 \bqo$,
while the $z$ fields
have a mass $\blam - 4 J_1 \bqo$ which vanishes at the
transition to the LRO
phase. The $\pi$ fields can therefore
be safely integrated out,
and ${\cal L}$ yields
the following effective action, valid at distances much
larger than the lattice
spacing~\cite{self2,self3}:
\begin{equation}
S_{eff} =
\int \frac{d^2 r}{\sqrt{8}a} \int_{0}^{c\beta}
d\ttau \left\{
|(\partial_{\mu} - iA_{\mu})z^{\alpha}|^2
+ \frac{\Delta^2}{c^2}
|z^{\alpha} |^2\right\},
\label{sefp}
\end{equation}
Here $\mu$ extends over $x,y,z$,
$c = \sqrt{8}J_1 \bqo a$ is the spin-wave velocity,
$\Delta = (\lambda^2 - 16J_1^2 \bqo^2 )^{1/2}$ is the gap
towards spinon excitations,
and $A_{\ttau} = A_{\tau}/c$.
Thus, in its final form, the long-wavelength theory consists
of a massive, spin-1/2, relativistic, boson $z^{\alpha}$ (spinon)
coupled to a compact $U(1)$ gauge field.

At distances larger than $c/\Delta$, we may safely integrate out the
massive $z$ quanta and obtain a a compact $U(1)$ gauge theory in 2+1
dimensions. This theory was argued by
Polyakov~\cite{polyakovbook,polyakov} to be permanently in a
confining phase, with the confinement driven by ``instanton''
tunnelling events. The compact $U(1)$ gauge force will therefore
confine the $z^{\alpha}$ quanta in pairs. In the present theory, the
confinementlength scale turns out to be exponentially large in $N$:
$\sim e^{c N}$ where the constant $c$ diverges logarithmically as
$\Delta \rightarrow 0$~\cite{ganpathy}. There are thus no free
spin-1/2 bosonic excitations for any finite $N$ and all low-lying
modes carry integral spin. The presence of an unbroken $U(1)$ gauge
force in the fluctuations has therefore completely disrupted the
simple mean-field structure of these states.
The spectrum of the $A_{\mu}$ quanta also acquires a gap from
instantons effects which is
exponentially small in $N$~\cite{polyakov,ganpathy}.

The properties of the $(0, \pi)$ phase are very similar to those of
the $(\pi , \pi)$ phase considered above, and will therefore not be
discussed here. It can be shown quite generally that any quantum
disordered state which has appreciable commensurate, collinear spin
correlations willl have similar properties: confined spinons and a
collective mode described by a compact $U(1)$ gauge theory.

We now turn to a study of the incommensurate phases. It is not
difficult to show that in this case it is not possible to satisfy the
constraints (\ref{lowenergy}) at any point in the Brillouin zone for
all the non-zero $Q_p$. This implies that there is no gapless
collective mode in the incommensurate SRO phases.
The structure of the theory is simplest in the vicinity of a
transition to a commensurate collinear phase:
we now examine the effective action as one moves
from the $(\pi , \pi)$-SRO phase into the $(q,q)$-SRO phase
(Figs~\ref{sp2f3}-\ref{sp2f4})
(a very similar analysis can be performed at the
boundary between the $(\pi , \pi)$-SRO and the $(\pi , q)$-
SRO phases). This transition is characterized by a
continuous
turning on of non-zero values of $Q_{i,i+\hy+\hx}$,
$Q_{i,i+2\hx}$ and $Q_{i,i+2\hy}$. It is easy to see from
Eqn (\ref{gaugetrans}) that these fields transform as scalars of
charge $\pm 2$ under the gauge transformation associated
with $A_{\mu}$. Performing a gradient expansion upon the
bosonic fields coupled to these scalars we find that the
Lagrangian ${\cal L}$ of the $(\pi , \pi)$-SRO phase gets
modified to
\begin{equation}
{\cal L} \rightarrow {\cal L} + \int \frac{d^2 r}{a} \left(
\vec{\Phi}_A \cdot \left(
\cj_{\alpha\beta}\psi_1^{\alpha}\vec{\nabla}\psi_1^{\beta}
\right) +
\vec{\Phi}_B \cdot \left(
\cj^{\alpha\beta}\psi_{2\alpha}\vec{\nabla}\psi_{2\beta}
\right) + \mbox{c.c.} \right)
\end{equation}
where $\vec{\Phi}_{A,B}$ are two-component scalars
$\equiv (J_3 Q_{3,x} + J_2 Q_{2,y+x},
J_3 Q_{3,y} + J_2 Q_{2,y+x} )$ with the sites on the ends of
the link
variables on sublattices $A,B$. Finally, as before, we
transform to the $z,\pi$ variables, integrate out the $\pi$
fluctuations and obtain~\cite{self5}
\begin{equation}
S_{eff} =
\int \frac{d^2 r}{\sqrt{8}a} \int_{0}^{c\beta}
d\ttau \left\{
|(\partial_{\mu} - iA_{\mu})z^{\alpha}|^2
+ r
|z^{\alpha} |^2 + \vec{\Phi} \cdot \left(
\cj_{\alpha\beta}z^{\alpha}\vec{\nabla} z^{\beta}\right) +
\mbox{c.c.}
+ V(\Phi) \right\} + \dots,
\label{hgs}
\end{equation}
Here $r= \Delta^2 /c^2$, $\vec{\Phi} = (\vec{\Phi}_A + \vec{\Phi}_B^{\ast} )
/(2J_1 \bqo a)$ is a
scalar of
charge $-2$; terms higher order in $\vec{\Phi}$ have been dropped.
We have also added a phenomenological potential
$V(\Phi )$ which is generated by short wavelength
fluctuations of the $b^{\alpha}$ quanta.
This effective action is also the simplest theory that can be written
down which couples a spin-1/2, charge 1, boson $z^{\alpha}$, a
compact $U(1)$ gauge field $A_{\mu}$, and a two spatial component,
charge $-2$, spinless boson $\vec{\Phi}$. It is the main result of
this section and summarizes essentially
all of the physics we are trying to describe.
We now describe the various phases of $S_{eff}$
\begin{enumerate}
\item
\underline{Commensurate, collinear, LRO:} $\langle z^{\alpha} \rangle
\neq 0$, $\langle \vec{\Phi} \rangle = 0$\\
This is the state with commensurate, collinear, magnetic LRO
\item
\underline{Commensurate, collinear, SRO:} $\langle z^{\alpha} \rangle
= 0$, $\langle \vec{\Phi} \rangle = 0$\\
This is the quantum-disordered state with collinear spin correlations
peaked at $(\pi , \pi)$. Its properties where described at length
above. The compact $U(1)$ gauge force confines the $z^{\alpha}$
quanta. The spinless collective mode associated with the gauge
fluctuations also has a gap.
\item
\underline{Incommensurate, coplanar, SRO:} $\langle z^{\alpha} \rangle
= 0$, $\langle \vec{\Phi} \rangle \neq 0$\\
This is the incommensurate phase with SRO at $(q,q)$ which we want
to study. It is easy to see that condensation of $\vec{\Phi}$
necessarily implies the appearance of incommensurate SRO:
ignore fluctuations of $\vec{\Phi}$ about $\langle \vec{\Phi} \rangle$
and diagonalize the quadratic form controlling the $z^{\alpha}$
fluctuations; the minimum of the
dispersion
of the $z^{\alpha}$ quanta is at a non-zero wavevector
\begin{equation}
\bk_0 =
(\langle\Phi_x\rangle
, \langle\Phi_y\rangle )/2
\end{equation}
The spin structure factor will therefore have a maximum at an
incommensurate wave\-vec\-tor. This phase also has a broken lattice
rotation symmetry due to the choice of orientation in the $x-y$ plane
made by $\vec{\Phi}$ condensate.
The condensation of $\vec{\Phi}$ also has a dramatic impact on the
nature of the force between the massive $z^{\alpha}$ quanta. Detailed
arguments have been presented by Fradkin and Shenker~\cite{fradshenk} that the
condensation of a doubly charged Higgs scalar quenches the confining
compact $U(1)$ gauge force in 2+1 dimensions between singly charged
particles.
Applied to the present problem, this implies that the charge $-2$
field $\vec{\Phi}$ condenses and deconfines the $z^{\alpha}$ quanta.
The excitation structure is therefore very similar to that of the
mean-field theory: spin-1/2, massive bosonic spinons and spinless
collective modes which have a gap. The collective mode gap is present
in this case even at $N=\infty$ and is associated with the
condensation of $\vec{\Phi}$.
\item
\underline{Incommensurate, coplanar, LRO:} $\langle z^{\alpha} \rangle
\neq 0$, $\langle \vec{\Phi} \rangle \neq 0$\\
The condensation of the $z$ quanta at the wavevector $\bk_0$ above
leads to incommensurate LRO, with the spin condensate spiraling in
the plane.
\end{enumerate}

\section*{\normalsize\sl 6.B.2 Triangular and Kagom\'{e} Lattices}
We will focus mainly on the nature of the fluctuations on the
triangular lattice~\cite{kagome}.
The properties of the kagom\'{e} lattice are very
similar and have been discussed elsewhere~\cite{kagome}.

The magnetically ordered state on the triangular lattice is coplanar.
 From the results on the $J_1 - J_2 - J_3$ model we anticipate that
the fluctuations about the quantum disodered state on the triangular
lattice will be similar to
those of the incommensurate, coplanar, SRO states - there will be no
gapless gauge modes and the spin-1/2 spinons remain unconfined.

We now present a few details verifying this conjecture.
It is not
difficult to see that the constraint (\ref{lowenergy}) for low-lying
gauge modes can be satisfied at most two of the values
of $p=1,2,3$ (corresponding to the $\hat{e}_p$ in Eqns.~(\ref{defei}))
at any point of the Brillouin zone. One such point is
the wavevector
\begin{equation}
{\bf g}_a = \frac{2\pi}{\sqrt{3} a} (0, 1)
\end{equation}
where
\begin{eqnarray}
g_{a1} & = & \pi \nonumber \\
g_{a2} & = & - \pi \nonumber \\
g_{a3} & = & 0
\end{eqnarray}
Taking the continuum limit with the fields varying with momenta with
close to ${\bf g}_a$ we find that the $I_{pq}$ in (\ref{ipqdef})
depend only upon
gradients of $\Theta_1$ and $\Theta_2$. Under gauge transformations
near the momentum ${\bf g}_a$, the bosons
\begin{equation}
b_i^{\alpha}~~\mbox{carry charges}~~ \exp ( i {\bf g}_a \cdot {\bf r}_i ).
\label{staggered}
\end{equation}
It can be verified that these
charges only take the values $\pm 1$ on the lattice sites. We have therefore
imposed a certain `staggering' of the charge assignments of the bosons
which is quite analogous to that in the square-lattice
antiferromagnets in Sec. 6.B.1.
It is also helpful to parametrize the $\Theta_p$ in the following
suggestive manner
\begin{eqnarray}
\Theta_1 ( {\bf r} ) & = & i A_{a1} ( {\bf r} ) e^{i {\bf g}_a \cdot {\bf r} }
\nonumber \\
\Theta_2 ( {\bf r} ) & = & - i A_{a2} ( {\bf r} ) e^{i {\bf g}_a \cdot {\bf r}
} \nonumber \\
\Theta_3 ( {\bf r} ) & = & \Phi_a ( {\bf r} ) e^{i {\bf g}_a \cdot {\bf r} }
\end{eqnarray}
It can be verified that the condition for the reality of $\Theta_p$
is equivalent to demanding that $A{a1}, A_{a2}, \Phi_a$ be real.
We will now take the continuum limit with $A{a1}, A_{a2}, \Phi_a$ varying
slowly on the scale of the lattice spacing. It is then not difficult to
show that the invariants $I_{pq}$ then reduce to (after a Fourier
transformation):
\begin{eqnarray}
I_{12} & = & \hat{e}_2 \cdot \vec{\nabla} A_{a1} -
\hat{e}_1 \cdot \vec{\nabla} A_{a2} \nonumber \\
I_{31} & = & \hat{e}_1 \cdot \vec{\nabla} \Phi_a - 2 A_{a1} \nonumber \\
I_{32} & = & \hat{e}_2 \cdot \vec{\nabla} \Phi_a - 2 A_{a2}
\end{eqnarray}
Thus the $A_{a1}, A_{a2}$ are the components of the connection of a
gauge symmetry
denoted $U_a (1)$; the components are taken along an `oblique' co-ordinate
system defined by the axes $\hat{e}_1 , \hat{e}_2$. The field
$\Phi_a$ transforms as the phase of charge 2 Higgs field under $U_a (1)$.

A very similar analysis can be carried out near the two other points in the
Brillouin zone where the other pairs of values of $\cos ( k_p /2) $ vanish.
These are the points
\begin{eqnarray}
{\bf g}_b &=& \frac{2\pi}{\sqrt{3} a}
\left(\frac{\sqrt{3}}{2}, -\frac{1}{2} \right) \nonumber \\
{\bf g}_c &=& \frac{2\pi}{\sqrt{3} a}
\left(\frac{-\sqrt{3}}{2}, -\frac{1}{2} \right)
\end{eqnarray}
which introduce the continuum symmetries $U_b (1)$ and $U_c (1)$
respectively. The $\Theta_p$ now reduce in the continuum limit
to fields $\Phi_b , A_{b2} , A_{b3}$ and
$A_{c1} , \Phi_c , A_{c3}$ respectively.
Thus in the continuum limit the lattice $U(1)$ gauge symmetry has been
replaced by a $U_a (1) \times U_b (1) \times U_c (1)$ gauge symmetry.
The three gauge symmetries correspond to the three different ways the
triangular lattice can be distorted into a rectangular lattice with diagonal
bonds: the phases on the horizontal and vertical bonds behave like
gauge connections while the phases on the diagonal bonds become charge 2 Higgs
fields.
The system also possesses spin-1/2 bose excitations which carry charges
$\pm 1$ of all 3 symmetries.

The condensation of all of the Higgs fields is implicit, and there are
therefore
no low-lying physical gauge excitations.
As in Sec. 6.B.1, we conclude that
the instantons
are
quenched and that unit charges are expected to be
unconfined~\cite{fradshenk}; in particular the spin-1/2
bose quanta, which
carry the $U_a (1)$ charges specified in (\ref{staggered}), and analogous
$U_b (1)$ and $U_c (1)$ charges,
will remain unconfined.

\section*{\normalsize\bf 6.C Berry Phases}

We found two different radically different consequences of gauge
fluctuations in the quantum-disordered phases in section 6.B:
\newline
({\em a\/}) the commensurate collinear phases appeared
to have gapless gauge modes; however, they acquired a gap from
instanton effects which also led to the confinement of spinons.
\newline
({\em b\/}) the coplanar phases on the square, triangular, and
kagom\'{e} phases had only gapped gauge fluctuations; the mean-field
spectrum was stable to fluctuation effects and the bosonic spinons
remained unconfined.
\newline
It was also clear from the analyses in Sections 6.B and 3 that the
Berry phases of small fluctuations about the mean-field saddle points
had been properly accounted for.

The only possibility that has not been explored carefuly is that the
topologically non-trivial gauge field configurations (instantons,
vortices) might
posses some non-trivial Berry phases. Such phases are of course not
present in a theory with a simple Maxwell action for the gauge-field
fluctuations; in the present theory the effective action is much more
complicated and the bosons propogate in a complicated gauge-field background
can acquire Berry phases. Such effects can, in principle, be present in
both the collinear and coplanar SRO phases. However, in the vicinity of the
transition to the LRO phases, instanton effects are not expected to
be dominant in the coplanar states; in contrast the collinear phases
have already been shown to be radically modified by instanton
effects.

\begin{figure}
\begin{picture}(400,200)
\multiput(60,25)(50,0){3}{\line(1,0){30}}
\multiput(60,75)(50,0){3}{\line(1,0){30}}
\multiput(60,125)(50,0){3}{\line(1,0){30}}
\multiput(60,175)(50,0){3}{\line(1,0){30}}
\multiput(50,35)(0,50){3}{\line(0,1){30}}
\multiput(100,35)(0,50){3}{\line(0,1){30}}
\multiput(150,35)(0,50){3}{\line(0,1){30}}
\multiput(200,35)(0,50){3}{\line(0,1){30}}
\put(50,25){\makebox(0,0){\large $B$}}
\put(100,25){\makebox(0,0){\large $A$}}
\put(150,25){\makebox(0,0){\large $B$}}
\put(200,25){\makebox(0,0){\large $A$}}
\put(50,125){\makebox(0,0){\large $B$}}
\put(100,125){\makebox(0,0){\large $A$}}
\put(150,125){\makebox(0,0){\large $B$}}
\put(200,125){\makebox(0,0){\large $A$}}
\put(50,75){\makebox(0,0){\large $A$}}
\put(100,75){\makebox(0,0){\large $B$}}
\put(150,75){\makebox(0,0){\large $A$}}
\put(200,75){\makebox(0,0){\large $B$}}
\put(50,175){\makebox(0,0){\large $A$}}
\put(100,175){\makebox(0,0){\large $B$}}
\put(150,175){\makebox(0,0){\large $A$}}
\put(200,175){\makebox(0,0){\large $B$}}
\put(75,50){\makebox(0,0){\large $W$}}
\put(75,150){\makebox(0,0){\large $W$}}
\put(175,50){\makebox(0,0){\large $W$}}
\put(175,150){\makebox(0,0){\large $W$}}
\put(125,100){\makebox(0,0){\large $Y$}}
\put(125,50){\makebox(0,0){\large $X$}}
\put(125,150){\makebox(0,0){\large $X$}}
\put(75,100){\makebox(0,0){\large $Z$}}
\put(175,100){\makebox(0,0){\large $Z$}}
\end{picture}
\caption{\sl The $A,B$ sublattices of the lattice of spins and the sublattices
$W,X,Y,Z$ of the dual lattice.}
\label{sublattices}
\end{figure}
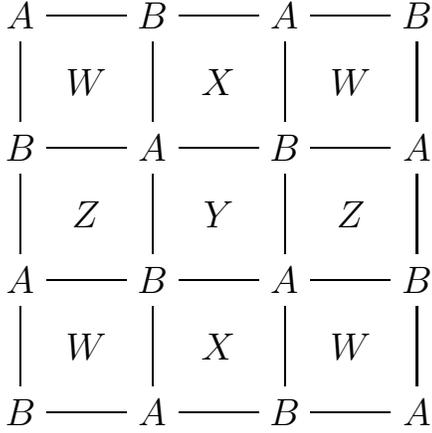

Detailed calculations of instanton Berry phase effects have been
performed in the collinear phases. These calculations are rather
involved and the reader is referred to Ref.~\cite{self3} for further
details: we will simply present the results here. It was first shown
by Haldane~\cite{hedge}, using the non-linear sigma
model formulation of Section
3 for fluctuations in the $(\pi, \pi)$ LRO state,
that `hedgehog' ${\bf n}$ configurations had the following Berry
phase term in its action
\begin{equation}
S_B = i \frac{\pi n_b}{2} \sum_s m_s \zeta_s
\label{berrypipi}
\end{equation}
where $n_b = 2S$, $m_s$ is the charge of an hedgehog centered on the
square lattice plaquette numbered $s$, and $\zeta_s$ takes the values
$0,1,2,3$ on the $W,X,Y,Z$ plaquette sublattices
(Fig~\ref{sublattices});
it can
be shown that no symmetry is broken by this choice of phases, and that
all observable correlation functions are invariant under the full lattice
symmetry group (in the absence of any dynamical symmetry breaking, of
course).
Note
that this phase is always unity for even-integer spins.

Subsequently, Refs~\cite{self2,self3} examined these Berry phases directly in
the commensurate, collinear SRO phases.
It was argued that the remnants of the hedgehogs in the LRO phases
were precisely the instantons in the compact $U(1)$ gauge field of
Section 6.B.
The Berry phase of bosons
propagating in an instanton gauge-field background was evaluated and
found to be identical to
(\ref{berrypipi}) in the $(\pi , \pi)$ SRO phase; $m_s$ is now the
total flux emannating from the instanton divided by $2\pi$. The result in the
$(0, \pi)$ SRO phase was different~\cite{self5}:
\begin{equation}
S_B = i \pi n_b \sum_s m_s \left[R_{sx}\right]
\end{equation}
where $\left[R_{sx}\right]$ is the integer part of $R_x$, the
$x$-coordinate of the plaquette $s$; this phase is unity for all
integer spins.
The result in the decoupled state was imply two copies of the $(\pi ,
\pi)$ SRO phase.

An analysis of the dynamics of the instantons was then carried
out~\cite{self2,wei}.
In the SRO phase the instantons interact with a Coulombic
$1/R$ potential; the instanton plasma can therefore be mapped
onto a dual
sine-Gordon model in which the instanton Berry phases appear
as frustrating phase-shifts in the arguments of the cosine
term. Finally analysis of this sine-Gordon model showed that non-unity
Berry phases led to spontaneous breaking of a lattice rotational
symmetry through the appearance of spin-Peierls order. The nature of
the spin-Peierls ordering depended strongly on the value of $\mnb$.

For the case of $(\pi , \pi)$ SRO the spin-Peierls ordering is shown
in Fig~\ref{sporder}.
The valence-bonds align in columns (4-fold degenerate) for half-integer spins,
and along lines (2-fold degenerate)
for integer spins. Only for even integer spins is
there no breaking of symmetry and we obtain a valence-bond-solid
state~\cite{aklt}.
\setlength{\unitlength}{0.8pt}
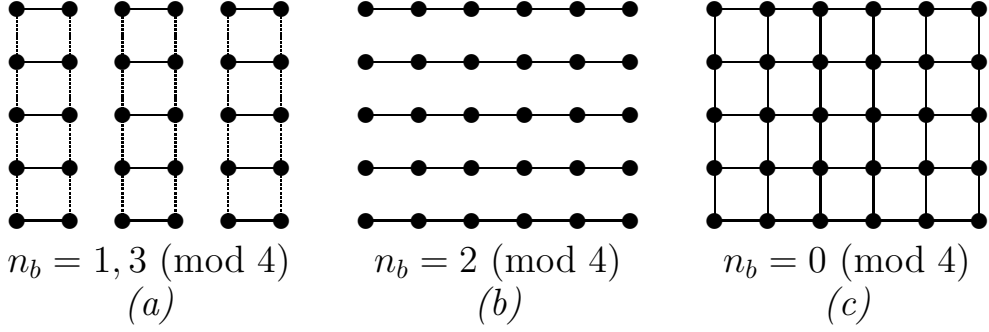
\begin{figure}
\begin{picture}(500,170)
\multiput(15,50)(0,25){5}{\circle*{7}}
\multiput(40,50)(0,25){5}{\circle*{7}}
\multiput(65,50)(0,25){5}{\circle*{7}}
\multiput(90,50)(0,25){5}{\circle*{7}}
\multiput(115,50)(0,25){5}{\circle*{7}}
\multiput(140,50)(0,25){5}{\circle*{7}}
\multiput(15,50)(0,25){5}{\line(1,0){25}}
\multiput(65,50)(0,25){5}{\line(1,0){25}}
\multiput(115,50)(0,25){5}{\line(1,0){25}}
\multiput(15,50)(50,0){3}{\dashbox{1}(25,100){}}
\multiput(180,50)(0,25){5}{\circle*{7}}
\multiput(205,50)(0,25){5}{\circle*{7}}
\multiput(230,50)(0,25){5}{\circle*{7}}
\multiput(255,50)(0,25){5}{\circle*{7}}
\multiput(280,50)(0,25){5}{\circle*{7}}
\multiput(305,50)(0,25){5}{\circle*{7}}
\multiput(180,50)(0,25){5}{\line(1,0){125}}
\multiput(345,50)(0,25){5}{\circle*{7}}
\multiput(370,50)(0,25){5}{\circle*{7}}
\multiput(395,50)(0,25){5}{\circle*{7}}
\multiput(420,50)(0,25){5}{\circle*{7}}
\multiput(445,50)(0,25){5}{\circle*{7}}
\multiput(470,50)(0,25){5}{\circle*{7}}
\multiput(345,50)(0,25){5}{\line(1,0){125}}
\multiput(345,50)(25,0){6}{\line(0,1){100}}
\put(77.5,30){\makebox(0,0){\large $n_b = 1,3$ (mod 4)}}
\put(242.5,30){\makebox(0,0){\large $n_b = 2$ (mod 4)}}
\put(407.5,30){\makebox(0,0){\large $n_b = 0$ (mod 4)}}
\put(77.5,10){\makebox(0,0){\large\em (a)}}
\put(242.5,10){\makebox(0,0){\large\em (b)}}
\put(407.5,10){\makebox(0,0){\large\em (c)}}
\end{picture}
\caption{\sl Symmetry of non-N\'{e}el ground states of
$H$
as a function of $\mnb$ with the minimum possible
degeneracies of 4,2,1 respectively ($n_b = 2S$
for $SU(2)$). The full-dotted-blank lines represent different values
of $\langle \hat{\bf S}_i \cdot \hat{\bf S}_j\rangle $ on the links.}
\label{sporder}
\end{figure}
It is interesting that an analysis of the quantum-dimer
model~\cite{dimernum}
for $n_b =1$ also found spin-Peierls order of the columnar type. Thus
the $N\rightarrow\infty$, $n_b=1$ theory agrees with the $N\rightarrow
\infty$, $n_b / N$ fixed, $n_b=1 \mf$ theory.

In the ``decoupled'' SRO phase, the above analysis applies
to each sublattice
separately, giving for $n_b = 1 \mf$
the type of spin-Peierls correlations shown in Fig~\ref{sp2f1}.
There is a total of $4\times 4 = 16$ states for this
case but coupling between the sublattices will
reduce this to 8 states, all of one of the two types shown.
The state with the `dimers' parallel to one another has
more possibilities for resonance using the $J_1$ bonds and
is likely to be the ground state.
For $n_b =
2 \mf$, there will be $2\times 2/2 = 2$ states, and for
$n_b = 0 \mf$, just one.

For the $(0, \pi)$ SRO state the
spin-Peierls order of the type shown in
Fig~\ref{sp2f1} for $n_b$ odd (half-integer spins), and a VBS state
for $n_b$ even (integer spins).  Combined with the
choice $(0,\pi)$ or $(\pi,0)$ this gives
degeneracies $2,4,2,4$ for $n_b=0,1,2,3 \mf$.

\section*{\normalsize\bf 6.D Summary}
The above analysis has been rather involved, but the essential results
are rather simple. Let us recall here the main properties of the
quantum disordered phases found in two dimensional antiferromagnets.
\begin{itemize}
\item
\underline{Commensurate, collinear phases}\\
The spin structure factor has a well-defined maximum at a commensurate point
in the Brillouin zone. The spinon excitations are massive and confined
in pairs - there are no low-lying excitations with half-integral
spins. There is a spinless collective mode which has a gap induced by
instanton effects. Spin-Peierls order is present and its nature
depends on the value of $\mnb$ (for the square lattice).
\item
\underline{Incommensurate, coplanar phases}\\
The spin structure factor has well-defined maxima at incommensurate points
in the Brillouin zone. The spin-1/2 spinon excitations are massive,
deconfined and carry bosonic statistics.
The spinless collective mode has a gap induced by the
condensation of a Higgs field: the magnitude of the condensate is
proportional to the incommensuration. There is a broken lattice
reflection symmetry associated with the choice of an axis about which
the fluctuating spiral order is present.
\item
\underline{Commensurate, coplanar fluids on the triangular and
kagom\'{e} lattices}\\
These phases from the incommensurate coplanar states only in that the
peak of the structure factor is at a commensurate point, and there is
no broken symmetry. These states thus appear to violate Laughlin's
fractional quantization principle~\cite{laughlin} that all spin-1/2
excitations of featureless spin-fluids should posses fractional
statistics: the spinons in the present theory are bosonic.
\end{itemize}

\section*{\normalsize\bf 6. Comparison with numerical and series
results}

Many
numerical~\cite{elbio} and series
analyses~\cite{rajiv}
have appeared on the $J_1-J_2-J_3$ model with $J_3 = 0$ for the spin-1/2
$SU(2)$ model {\em i.e.} $N=1$, $n_b =1$ in the notation of
this paper. They find $(\pi , \pi )$-LRO at small $J_2 /
J_1$, $(\pi , 0)$-LRO at large $J_2 / J_1$ and an
intermediate SRO phase around $J_2 / J_1 = 1/2$. This is in
agreement (see Fig~\ref{sp2f1})
with our prediction in Section 6.B that the
phase boundary between $(\pi , \pi )$-LRO and $(\pi , \pi)$-
SRO bends downwards at finite $N$ with increasing $J_2 /
J_1$. Analyses of this intermediate
phase at $J_2 / J_1$~\cite{elbio,rajiv} shows clear evidence
of columnar spin-Peierls ordering~\cite{self2}, also in agreement
with the results of Section 6.C.
An additional intermediate phase with $(0, \pi)$-SRO
has not been ruled out.

There has also been intensive work recently on the ground state of the
$SU(2)$ antiferromagnet on the triangular lattice. For spin-1/2, there
is good evidence~\cite{zengelser,rajivkag} that the ground state is
quantum disordered. There are preliminary indications~\cite{chalker} that
spin-Peierls order is absent in this state (which was predicted in
Ref.~\cite{kagome} and discussed in Section 6.B.2), although larger
system sizes are required before any firm conclusion can be reached.
Another interesting issue of the kagom\'{e} is the nature of the
large-spin magnetically ordered ground state selected by quantum
fluctuations. A number of investigators~\cite{chubkag,huserut,kallin,shender}
have found the $\sqrt{3} \times \sqrt{3}$ discussed in Section 6.A.2.

\section*{\normalsize\bf 7. CONCLUSIONS}

It should be clear from all the discussion in this paper that there
has been a great deal of progress in our understanding of non-random
two dimensional antiferromagnets. The magnetically ordered states had been
understood many years ago by semiclassical analyses. This course
therefore focussed on recent work on the nature of the quantum phase transition
to the quantum disordered phase and the properties of the quantum
disordered phase itself. A partial classification of the different types of
possible quantum disordered states has emerged, and was summarized in Section
6.D - the reader is urged to review this section, even if he/she did not
have the patience to read the rest of Section 6. Also interesting was the
fundamental connection between the properties of the quantum disordered
states and the nature of the spin ordering in the magnetically ordered states.

Comparison of these results with experiments is complicated by the
ubiquitous presence of randomness. Unlike classical systems, even weak
randomness
has a strong effect on the properties of quantum phase transitions and
the low-energy properties of the quantum-disordered phase. In the
renormalization
group language, randomness is almost always a relevant perturbation at
fixed points controlling the properties of quantum phase transitions
and quantum-disordered phase.
Our comparision with experiments has therefore been limited to a simple
scaling analysis of phase transitions in random quantum antiferromagnets -
this was summarized in Section 5.
More detailed study of random quantum antiferromagnets is therefore clearly
a high priority for future work. Recently, there has been some progress in
solving random quantum magnets with infinite-range
exchange interactions~\cite{qsg,qlrm}; a variety of unusual results were
obtained, including the presence of gapless excitations in the
quantum disordered phase.
There have also been exact solutions of random quantum spin chains which
are very instructive~\cite{fisher}. One hopes that these works are just the
first stages of much further work on random quantum spin systems: the outlook
for more interesting results is promising and the number of available
experimental systems continues to grow.

\section*{\normalsize\bf Acknowledgements}
I thank N. Read and Jinwu Ye for several fruitful collaborations which led
to the work reviewed here.
This research was
supported by NSF Grant No. DMR-8857228
and by a fellowship from the A.P. Sloan Foundation.

\end{document}